\documentclass[twocolumn]{aastex631}

\newif\ifrev
\revfalse
\definecolor{darkred}{rgb}{0.75, 0, 0}
\definecolor{darkblue}{rgb}{0, 0, 1}
\newcommand{\rev}[1]{\ifrev\leavevmode{\bf#1}\else #1\fi}

\newif\ifrevv
\revvfalse

\newif\ifrevvv
\revvvfalse
\definecolor{darkred}{rgb}{0.75, 0, 0}
\definecolor{darkblue}{rgb}{0, 0, 1}
\newcommand{\revv}[1]{\ifrevv\leavevmode{\textbf{#1}}\else #1\fi}
\newcommand{\revvv}[1]{\ifrevvv\leavevmode{\bf#1}\else #1\fi}

\begin{document}



\title{\rev{Asteroseismic Inversions for Internal Sound Speed Profiles of Main-sequence Stars with Radiative Cores} }

\correspondingauthor{Lynn Buchele}
\email{lynn.buchele@h-its.org} 

\author[0000-0003-1666-4787]{Lynn Buchele} 
\affiliation{Heidelberger Institut für Theoretische Studien, Schloss-Wolfsbrunnenweg 35, 69118 Heidelberg, Germany}
\affiliation{Center for Astronomy (ZAH/LSW), Heidelberg University, Königstuhl 12, 69117 Heidelberg, Germany}

\author[0000-0003-4456-4863]{Earl P.~Bellinger}
\affiliation{Max-Planck-Institut für Astrophysik, Karl-Schwarzschild-Straße 1, 85748 Garching, Germany}
\affil{Department of Astronomy, Yale University, New Haven, CT 06520, USA}
\affiliation{Stellar Astrophysics Centre, Department of Physics and Astronomy, Aarhus University, Ny Munkegade 120, 8000 Aarhus C, Denmark}

\author[0000-0002-1463-726X]{Saskia Hekker}
\affiliation{Heidelberger Institut für Theoretische Studien, Schloss-Wolfsbrunnenweg 35, 69118 Heidelberg, Germany}
\affiliation{Center for Astronomy (ZAH/LSW), Heidelberg University, Königstuhl 12, 69117 Heidelberg, Germany}

\author[0000-0002-6163-3472]{Sarbani Basu}
\affil{Department of Astronomy, Yale University, New Haven, CT 06520, USA}

\author[0000-0002-4773-1017]{Warrick Ball}
\affil{Advanced Research Computing, University of Birmingham, Edgbaston, Birmingham B15 2TT, UK} 
\affil{School of Physics \& Astronomy, University of Birmingham, Edgbaston, Birmingham B15 2TT, UK} 

\author[0000-0001-5137-0966]{Jørgen Christensen-Dalsgaard}
\affiliation{Stellar Astrophysics Centre, Department of Physics and Astronomy, Aarhus University, Ny Munkegade 120, 8000 Aarhus C, Denmark}

\begin{abstract}

The theoretical oscillation frequencies of even the best asteroseismic models of solar-like oscillators show significant differences from observed oscillation frequencies. Structure inversions seek to use these frequency differences to infer the underlying differences in stellar structure.  While used extensively to study the Sun, structure inversion results for other stars have so far been limited. Applying sound-speed inversions to more stars allows us to probe stellar theory over a larger range of conditions, as well as look for overall patterns that may hint at deficits in our current understanding. To that end, we present structure inversion results for 12 main-sequence \rev{solar-type stars} \revv{with masses between 1~M\(_\odot\) and 1.15~M\(_\odot\)}. Our inversions are able to infer differences in the isothermal sound speed in the innermost 30\% by radius of our target stars. \revv{In half of our target stars, the structure of our best-fit model fully agrees with the observations. In the remainder, the inversions reveal significant differences between the sound-speed profile of the star and that of the model. We find five stars where the sound speed in the core of our stellar models is too low and one star showing the opposite behavior.} \rev{ For the two stars which our inversions reveal the most significant differences, we examine whether changing the microphysics of our models improves them and find that changes to nuclear reaction rates or core opacities can reduce, but do not fully resolve, the differences. }

\end{abstract}

\section{Introduction} \label{sec:intro}
The combination of high-precision photometric time series data from \emph{Kepler} \citep{2010Sci...327..977B}, astrometric parallax data from Gaia \citep{2016A&A...595A...1G}, and high-resolution spectroscopic measurements of \rev{effective temperatures and metallicities} \citep[for example from the \emph{Kepler} Follow-up Program,][]{ 2018ApJ...861..149F} provides an opportunity to test stellar evolution theory at unprecedented precision.

In particular, asteroseismology, which uses oscillation frequencies obtained from analysis of stellar light curves, provides a direct way to test the physics of stellar interiors \citep{2010aste.book.....A, 2017asda.book.....B}. This is possible because the star’s oscillation frequencies are sensitive to the internal structure of the star. By constructing stellar evolution models that seek to reproduce a star’s observed oscillation frequencies and surface properties (for example, luminosity, effective temperature, and metallicity), asteroseismology can be used to study a broad range of physics, including atomic diffusion, rotation, magnetic fields, and convection \citep[for an overview, see, e.g.,][]{2013ARA&A..51..353C, 2019LRSP...16....4G}. 
These asteroseismic models can be found using a variety of techniques, including \rev{Bayesian inference (e.g., \citealt{2015MNRAS.452.2127S,2017ApJ...835..173S, 2022MNRAS.509.4344A}), MCMC (e.g., \citealt{2008MmSAI..79..660B, 2012ApJ...749..109G, 2013MNRAS.435..242G, 2019MNRAS.484..771R, 2019ApJ...887L...1B, 2021RAA....21..226J}), machine learning (e.g., \citealt{2016ApJ...830...31B, 2019A&A...622A.130B, 2020MNRAS.491.4752B, 2020MNRAS.493.4987A, 2020MNRAS.499.2445H, 2023A&C....4200686G}), genetic algorithms \revv{(e.g., \citealt{2003JCoPh.185..176M, 2005A&A...437..575C, 2009ApJ...699..373M, 2014ApJS..214...27M, 2023RNAAS...7..164M})}, and Levenberg-Marquardt algorithms (e.g., \citealt{2002A&A...394L...5F, 2003Ap&SS.284..233T, 2005A&A...441..615M}).}

However, for stars with the highest quality asteroseismic data,  there are still discrepancies between models and observations. This tension between theoretical and observed frequencies suggests that our models need to be improved, although it does not directly suggest what those improvements should be. 

\rev{We aim to gain insight into the potential underlying structural differences between stellar models and observations using the technique of asteroseismic structure inversions. This technique uses the differences between the frequencies of an observed star and its model to infer localized information about the structure differences} \revv{(see e.g. \citealt{1991sia..book..519G, 1993afd..conf..399G, 2006mha..book.....P, 2020ASSP...57..171B, 2022FrASS...9.2373B}).}

In the case of the Sun, structure inversions have been used to study the equation of state, diffusion of heavier elements, and nuclear reaction rates in connection to the solar neutrino problem, \citep[for a review see, for example,][]{2016LRSP...13....2B, 2021LRSP...18....2C}. The high precision and large number of modes observed for the Sun allow structure inversions to probe a large extent of the solar interior, from \(0.06\,\rm{R}_\odot\) to \(0.96\,\rm{R}_\odot\). This, however, is not the case for other stars. \rev{Current asteroseismic observations are typically limited to modes of spherical degree $l=0,1,2$, with a few $l=3$ modes being observed in the best target stars. This limits  the range that can be probed with \revv{local structure inversions} to the near-core region, fractional radii between $\sim 0.05$ and $0.35$ \citep{2020ASSP...57..171B}.}

Nevertheless, there are several examples of structure inversions performed on stars other than the Sun, including the solar analogues 16~Cyg~A and 16~Cyg~B \citep{2017ApJ...851...80B,2022A&A...661A.143B}, a main-sequence star with a convective core \citep{2019ApJ...885..143B}, and a few subgiants with mixed modes \citep{2020IAUS..354..107K, 2021ApJ...915..100B}. \rev{Inversion techniques are also being developed for more massive stars \citep{2023A&A...675A..17V} and more evolved stars \citep{2018Natur.554...73G}.}
By looking at a larger number of stars, we can test the theory of stellar structure and evolution under a broader range of conditions, such as different masses, metallicities, ages, and evolutionary stages. Examining several stars at once also provides the opportunity to look for overall trends that may hint at deficits in our current understanding of stars. In this work, we focus on studying the most solar-like stars---main-sequence stars with radiative cores---using structure inversions.

\section{\rev{Forward Modeling}} \label{sec:modeling} 
The goal of a structure inversion is to infer the differences between the actual stellar structure and that of a reference model. \rev{As the structure inversion equation is based on a linear perturbation approach, the reference model must be suitably close to the actual star. Hence, we typically use the best-fit model obtained with some modeling procedure called forward modeling. Here, we describe the forward modeling procedure used to obtain our reference model for each target star. }
We created a grid of 16384 tracks using the r22.05.01 version of the stellar evolution code MESA \citep[][]{Paxton2011, Paxton2013, Paxton2015, Paxton2018, Paxton2019, Jermyn2022}. \rev{We vary the initial mass, initial helium mass fraction, metallicity, and mixing-length parameter  using a Sobol sequence \citep[see Appendix~B of][]{2016ApJ...830...31B, SOBOL196786}. Table~\ref{tab:grid_param} gives the range that was covered in each parameter.} All models in this grid use metal abundances scaled to the GS98 solar composition \citep{1998SSRv...85..161G}, and the corresponding opacity tables from OPAL \citep{Iglesias1993,
Iglesias1996} in the high-temperature range, and \citet{Ferguson2005} in the low-temperature range. The equation of state data are calculated with the MESA default blend of OPAL \citep{Rogers2002}, SCVH \citep{Saumon1995}, FreeEOS \citep{Irwin2004}, and Skye \citep{Jermyn2021}. For details of how this blending is handled, see \citet{Jermyn2022}. We use the \verb|pp_cno_extras_o18_ne22.net| reaction network and take our reaction rates from JINA REACLIB \citep{Cyburt2010} and  NACRE \citep{Angulo1999},  with additional tabulated weak reaction rates \citep{Fuller1985, Oda1994, Langanke2000}.  Electron screening is included via the prescription of \citet{Chugunov2007}.
Thermal neutrino loss rates are from \citet{Itoh1996}. Convection in the models is described using the time-dependent local convection formalism of \citet{1986A&A...160..116K}, which in the limit of long time steps reduces to standard mixing length theory as described in \citet{1968pss..book.....C}. The implementation details are given in \citet{Jermyn2022}. 
We account for atomic diffusion through gravitational settling, as described in \citet{Paxton2011}. \revv{We use an Eddington-gray atmosphere and include the structure of the atmosphere out to an optical depth of \(\tau = 10^{-3}\) when calculating both our oscillation frequencies and structure kernels.} The \rev{adiabatic} frequencies of the models were computed using GYRE \citep{Townsend2013, Townsend2018}.

\begin{deluxetable}{lcc}

\tablecaption{Grid Parameters} 
\label{tab:grid_param} 
\tablehead{\colhead{Parameter} & \colhead{Minimum Value} & \colhead{Maximum Value}} 
\startdata
$M [\rm{M}_{\odot}]$ & 0.7 &  1.2 \\
$Y_{\rm{initial}}$ & 0.24 &  0.29 \\
$Z_{\rm{initial}}$ & 0.0005  &  0.07 \\
$\alpha_{\rm{mlt}}$ & 1.3 & 2.4 \\
\enddata
\end{deluxetable}

For each target star, we find reference models by fitting the observed frequencies, effective temperature, and metallicity of each star. We take our frequency data from the \emph{Kepler} LEGACY sample \citep{2017ApJ...835..172L} and the \emph{Kepler} ages (KAGES) sample \citep{2016MNRAS.456.2183D}. In the case of 16~Cyg~A and 16~Cyg~B we use the frequencies given in \citet{2017A&A...604A..42R} labeled as Roxburgh(Davies). Spectroscopic measurements of the effective temperature and metallicity are from the \rev{combined stellar parameters reported by the \emph{Kepler} Follow-Up program \citep[][their Table~9]{2018ApJ...861..149F}. These values are computed by combining the results of four different spectroscopic analysis pipelines. We also adopt their suggested uncertainties of 100~K, 0.1~
dex for \(T_{\rm{eff}}\) and [Fe/H] respectively. The observational parameters we consider for each star are listed in Appendix~\ref{appendix:ref_models}.} \rev{For each target star, we search our grid to find the model that  minimizes}
\rev{
\begin{equation}
\chi^2_{\rm{fit}} = \chi^2_{\nu} + \chi^2_{T_{\rm{eff}}} + \chi^2_{[\rm{Fe/H}]} 
\label{equ:fit_chi2}
\end{equation} }
\rev{
where}
 \rev{
\begin{equation} 
\chi^2_{\nu} = \frac{1}{N} \sum_{i}^N \frac{(\nu_{i, \rm{obs}} - \nu_{i, \rm{mod}})^2}{\sigma_{\nu, i}^2},
\label{equ:freq_chi2} 
\end{equation} }
\rev{
\begin{equation} 
\chi^2_{T_{\rm{eff}}} = \frac{(T_{\rm{eff, obs}} - T_{\rm{eff, mod}})^2}{\sigma_{T_{\rm{eff}}}^2}, 
\label{equ:Teff_chi2} 
\end{equation} }

 \rev{and }
 \rev{
 \begin{equation} 
 \chi^2_{[\rm{Fe/H}]} = \frac{[\left(\rm{Fe/H}]_{\rm{obs}} - [\rm{Fe/H}]_{\rm{mod}}\right)^2}{\sigma_{[\rm{Fe/H}]}^2}. 
 \label{equ:FeH_chi2} 
 \end{equation} }
\rev{Here $N$ is the number of frequencies and the subscripts `obs' and `mod' refer to the observations and the model respectively. The model frequencies used to calculate \(\chi^2_{\nu}\) are first corrected for surface effects using the two-term correction from \citet{2014A&A...568A.123B}.}

\rev{ We scan our grid to find the parameters (\(M, Y_{\rm{initial}}, Z_{\rm{initial}}, \alpha_{\rm{mlt}}, X_{c}\)) that minimize \(\chi^2_{\rm{fit}}\). In the process, we interpolate in central hydrogen abundance along each track, \revv{but not we do not interpolate between the tracks}. In order to reduce the computational time necessary to find a best-fit model, we consider for subsequent analysis only models that are within \(6 \sigma\) of the effective temperature and metallicity values, as well as within \(10\sigma\) of the FLAME luminosity value from Gaia DR3 \citep{2016A&A...595A...1G, 2022arXiv220800211G, 2023A&A...674A..26C}. We then use these parameters, \revv{given in Appendix~\ref{appendix:ref_models},} to compute the reference model that will be used for structure inversions of each star.} We have made the FGONG structure files of our reference models as well as the inlists used to generate them publicly available. \footnote{\url{https://zenodo.org/records/10391300}}

\section{Inversions}\label{sec:inversions}
With a suitable reference model, we aim to use the differences between the frequencies of an observed star and the frequencies of the reference model to infer the underlying structure differences. We do this through the use of stellar structure kernels, which express the sensitivity of an oscillation mode frequency to a small perturbation to the structure. Mathematically, this is expressed in the kernel equation \rev{\citep{1990MNRAS.244..542D}}: 
\begin{equation} \label{equ:kern} 
\frac{\delta \nu_i} {\nu_i} = \int K_i^{(f_1, f_2)}(r) \frac{\delta f_1}{f_1} \: dr + \int K_i^{(f_2, f_1)}(r) \frac{\delta f_2}{f_2} dr \rev{+ \mathcal{O}(\delta^2)}
\end{equation} 
\rev{Here \(i\) is the index of the mode which corresponds to a specific pair of radial order ($n$) and spherical degree ($l$), \(\delta \nu{_i} / \nu{_i}  = (\nu_{i,\rm{obs}} - \nu_{i,\rm{mod}})/\nu_{i,\rm{mod}} \) is the relative difference  in frequency between the observed mode ($\nu_{i, \rm{obs}}$) and the corresponding mode of the reference model ($\nu_{i,\rm{mod}}$),  \(f_1\) and \(f_2\) are the stellar structure variables being considered, and \(K_i\) are the kernel functions of each mode.}
\rev{The mode kernel functions (\(K_i\)) are obtained though a linear perturbation of the oscillation equation (for more details, see \citealt{1991sia..book..519G}, \citealt{1999JCoAM.109....1K}, or \citealt{2002ESASP.485...95T}).} \rev{Initially, mode kernels were derived in terms of the squared sound speed \(c^2\) and density \(\rho\) \citep{1990MNRAS.244..542D}. From this expression, mode kernels for other pairs of variables have been derived, including for density and the first adiabatic exponent \(\Gamma_{1}\) \revv{(e.g., \citealt{1991sia..book..519G,1993afd..conf..399G}),  isothermal sound speed \(u = c^2/ \Gamma_{1}\) and helium mass fraction \(Y\) (e.g., \citealt{1997A&A...322L...5B, 1999JCoAM.109....1K,2015A&A...583A..62B,2017A&A...598A..21B}), and convective stability parameter $A$ and $\Gamma_1$ (e.g., \citealt{1996MNRAS.280.1244E, 1999JCoAM.109....1K, 2017A&A...598A..21B}).} For more details on changing the structure variable pair, see \citet{2011LNP...832....3K, 2017A&A...598A..21B}. Equation \ref{equ:kern} can also include a term that corrects for the surface effect; however, we instead correct for this in the calculation of the frequency differences. For the remainder of this paper, when we discuss model frequencies, we refer to the frequencies that have been corrected for surface effects.}

Each oscillation mode is sensitive to many points within the star, so to obtain localized information we implement the method of optimally localized averages (OLA, \revv{\citealt{1968GeoJ...16..169B, 1970RSPTA.266..123B}}) which uses a linear combination of the frequency differences. \rev{Neglecting second-order effects}, Equation~\ref{equ:kern} becomes 
\begin{equation} 
\sum_{i=1,N} \rev{c_{i} \frac{\delta \nu_{i}}{\nu_{i}}} = \int \mathcal{K}(r) \frac{\delta f_1}{f_1} \: dr + \int \mathcal{C}(r) \frac{\delta f_2}{f_2} \: dr.
\label{equ:lincomb} 
\end{equation} 
Here \(\mathcal{K}\) is the averaging kernel and  \(\mathcal{C}\) is the cross-term kernel.
These are constructed using a set of inversion coefficients \(c_i\): 
\begin{equation}
\mathcal{K} = \sum_{i=1,N} c_i K_i^{(f_1, f_2)} \quad {\rm{and}} \quad \mathcal{C} = \sum_{i=1,N} c_i K_i^{(f_2, f_1)} . 
\label{equ:K_def]} 
\end{equation} 
If \(\mathcal{K}\) is normalized to 1 and \(\mathcal{C}\) is small, then Equation~\ref{equ:avg_kern} reduces to 
\begin{equation} 
\label{equ:avg_kern} 
\sum_{i=1,N} \rev{c_{i} \frac{\delta \nu_{i}}{\nu_{i}}} \approx \int \mathcal{K}(r) \frac{\delta f_1}{f_1} \: dr  \approx \left<\frac{\delta f_1}{f_1} \right>,
\end{equation} 
and \(\mathcal{K}\) can be interpreted as the weight function of a mean over the structure difference \(\delta f_1/ f_1\). This is why \(\mathcal{K}\) is called the averaging kernel. In other words, if the coefficients \(c_i\) are chosen such that \(\mathcal{K}\) has most of its amplitude around a single target radius, \(r_{0}\), then the same linear combination of frequency differences provides a localized average difference of the structure variable \(f_1\) at that target radius. 

\subsection{\rev{Localized averaging kernels}}

\rev{To construct a localized averaging kernel, we use the method of multiplicative optimally localized averages (MOLA). For a MOLA inversion, we define a weight function, \(J = (r-r_{0})^{2}\),  that penalizes any amplitude of the averaging kernel away from the target radius.} 
In addition to the target radius, there are two \rev{trade-off parameters} that must be chosen: the error suppression parameter, \(\mu\), and the cross-term suppression parameter, \(\beta\). The inversion coefficients are found by minimizing

\begin{eqnarray} \label{equ:MOLA_min} 
 \int \left(\sum_{i=1,N} c_i K_i^{(f_1,f_2)} \right)^2 J(r_0,r) dr \\ \nonumber + \beta \int \left(\sum_{i=1,N} c_i K_i^{(f_2, f_1)}\right)^2 dr  
+ \rev{\mu \sum_{i=1,N} c_i c_j E_{ij},} 
\end{eqnarray} 
\rev{where \(E_{ij}\) are the elements of the error-covariance matrix.} \revv{Strictly speaking, there is another trade-off parameter in this formulation, the normalization of \(J\), which we have set to 1, while other implementations often use 12. However, this only changes the relative weight of the first term in Equation~\ref{equ:MOLA_min} and can be counteracted by changing \(\mu\) or \(\beta\). Thus, while the optimal values of \(\mu\) and \(\beta\) vary with the normalization of \(J\), the inversion results do not.}

\rev{The other standard method of constructing an averaging kernel is a variant of MOLA known as the method of subtractive optimally localized averages (SOLA, \revv{\citealt{1992A&A...262L..33P,1994A&A...281..231P}}). Previous works studying 16~Cyg~A and 16~Cyg~B have used SOLA \citep{2017ApJ...851...80B, 2022A&A...661A.143B}. We choose to use MOLA because it requires setting only two free parameters, as opposed to the three required for a SOLA inversion. Additionally, we find that MOLA is better able to suppress the amplitude of the averaging kernel at the surface.  For details on the differences between MOLA and SOLA see \citet[][Chapter~10]{2017asda.book.....B}.}

The next important consideration for a structure inversion is which pair of structure variables to use.  We use the \((u,Y)\) pair because the $Y$ kernels have low amplitude everywhere except in helium ionization zones \citep{2003Ap&SS.284..153B}, \rev{which naturally suppresses the cross-term kernel at the radii we are targeting.} The trade-off to this approach is that using \(Y\) as a structure variable requires the assumption of an equation of state. In the solar case, the error introduced by this assumption is significant in comparison to the other sources of uncertainty \citep{1997A&A...322L...5B}; however, due to the larger uncertainties on asteroseismic frequencies this is not the case for stars other than the Sun. Using $Y$ as a structure variable requires calculating several partial derivatives of \(\Gamma_1\). To be consistent with the blend of equation of state tables used in MESA, we \rev{obtained} these directly from MESA's equation of state module.

\subsection{\rev{Trade-off parameters}}
\rev{As Equation~\ref{equ:MOLA_min} shows, there are two trade-off parameters that must be chosen in the course of a structure inversion. The parameter \(\mu\) controls the balance between a well-localized averaging kernel and the amplification of uncertainties. } To choose an appropriate value of \(\mu\) for each inversion, we utilize a set of calibration proxy models. These models are found using the process described in Section \ref{sec:modeling}; however, they have slightly higher \(\chi^2_{\rm{fit}}\) values than the reference model. Since the structure of these models is known exactly, they can be used to determine how well the inversion recovers the underlying differences. \revv{We provide the details of this process in Appendix~\ref{appendix:mu}. }\revv{
Before accepting our inversion results, we visually inspect the averaging and cross-term kernels of all target radii for each star to ensure that the averaging kernels are well localized and the overall amplitude of the cross-term kernels are low.}

\subsection{\rev{Stellar mass \& radius}} \label{sect:MR} 
Another complication for inversions of stars other than the Sun is the lack of precise measurements of the stellar mass and radius. Since the frequencies of a star scale with its mean density, a mismatch in the mean density will lead to an offset in the inversion results \citep{2003Ap&SS.284..153B}. To minimize this, we invert for the relative difference in \(\hat{u} = u R/GM\), where \(R\) and \(M\) are the star's radius and mass, respectively, and $G$ is the gravitational constant. This is done by using mode kernels computed in a dimensionless form and using the dimensionless frequency differences. Previously, \citet{2021ApJ...915..100B} used dimensionless frequency differences calculated by subtracting off the weighted mean of the frequency differences \rev{(for details, see \citealt{2003Ap&SS.284..153B}).}

We have found that this method results in correct dimensionless frequencies only when the frequency differences caused by an incorrect mean density are larger than the differences introduced by the structural variation. Whether this is true cannot be determined purely by comparing the observed and modeled frequencies. Instead, we use a new method of calculating the dimensionless frequency differences using the dependence of the large frequency separation \rev{\((\Delta \nu)\)} on the mean density. \rev{The large frequency separation is the mean frequency difference between successive radial modes and is a proxy for the root mean density of the star} \revv{\citep{1967AZh....44..786V}}. The details of this method can be found in Appendix~\ref{appendix:mean density}. 

\rev{We calculate our value of \(\Delta \nu\) by taking the slope of a linear fit to the relationship between the \(l=0\) modes and their respective radial orders and use this to calculate the dimensionless frequency differences. These corrections mean that the uncertainty of our frequency differences are no longer independent, and hence, the error co-variance matrix \(E\), used in Equation~\ref{equ:MOLA_min} is not diagonal. We calculate it using a Monte Carlo simulation where each frequency is perturbed, 10\,000 times with Gaussian noise according to their measured uncertainties. These perturbations are applied before the frequencies are corrected for the surface effect, and so this procedure also accounts for the error correlation introduced by the surface term correction.  This same set of perturbed frequencies is then used to calculate the final inversion results.} We take the average of this distribution as our final inversion result and report the standard deviation as the uncertainty. 

    \revvv{This method of uncertainty estimation occurs after both the reference model and inversion parameters have been selected, and only propagates uncertainties due to the underlying frequencies. It is the same as the traditional definition of inversion uncertainties (e.g., Equation 4 of \citealt{2019ApJ...885..143B}) except that it accounts for correlation introduced during the pre-processing of our frequency differences.}

To validate both our method of finding reference models and our inversion results, we also obtain a reference model and inversion results using solar data that have been degraded to the level that was expected of results from \emph{Kepler}. These results are given in Appendix~\ref{appendix:mod_S}. 

\subsection{\rev{Overall inversion significance} }
For each star, we attempt structure inversions at six target radii, $r_{0}/R = 0.05, 0.10, 0.15, 0.20, 0.25, 0.30$, although in some cases we are only able to find suitable averaging kernels at five target radii. To quantify the disagreement between each target star and its model, across all target radii, we calculate a \(\chi^{2}_{\rm{inv}}\) as follows. For each star, there is a set of inversion results and their associated uncertainties \(v_j \pm u_j\). Since all the target radii use the same underlying data, their errors are correlated. The correlation between two target radii, \(r_j, r_k\), is \citep{2017asda.book.....B}. 
\begin{equation}
E_{r_j,r_k} = \frac{\sum_i c_i(r_j) \cdot c_i(r_k) \sigma_i^2}{\left[\sum_i c_i^2(r_j) \sigma_i^2\right]^{1/2} \left[\sum_i c_i^2(r_k) \sigma_i^2\right]^{1/2}}
\label{equ:err_corr}
\end{equation}
where \(c_i(r_j)\) is the inversion coefficient of the \(i\)-th mode for the \(j\)-th target radius and \(\sigma_i\) is the relative uncertainty of the \(i\)th mode frequency. The error correlation matrix, \(\mathbf{E}\), is the matrix with components \(E_{r_j, r_k}\) between all different target radii. 
The covariance matrix then is,
\begin{equation}
    \mathbf{C} = \mathbf{U}^{\rm{T}} \mathbf{E} \mathbf{U}
    \label{equ:covar}
\end{equation}
where \(\mathbf{U}\) is a diagonal matrix with the uncertainty of the inversion result for each target radius. Then  
\begin{equation}
    \chi_{\rm{inv}}^2 = \mathbf{V}^{\rm{T}}  \mathbf{C}^{-1}  \mathbf{V} 
    \label{equ:chi2_inv} 
\end{equation}
where \(\mathbf{V}\) is the vector of inversion results at each target radius. This \(\chi^{2}_{\rm{inv}}\) summarizes the overall significance of the inversion results for each star across all target radii, with larger values indicating larger disagreement.

In summary, after finding a reference model, we calculate the surface-term-corrected dimensionless frequency differences between the target star and the model. We then use our set of calibration models to choose \(\mu\) at each target radius and obtain our set of averaging kernels. With this, we use the inversion coefficients and the frequency differences to obtain our inferred localized differences in \(\hat{u}\) between the observed star and our reference model.

\section{Results and Discussion} \label{sec:results}

\begin{figure*}
    \centering
    \epsscale{1.1}

    \plotone{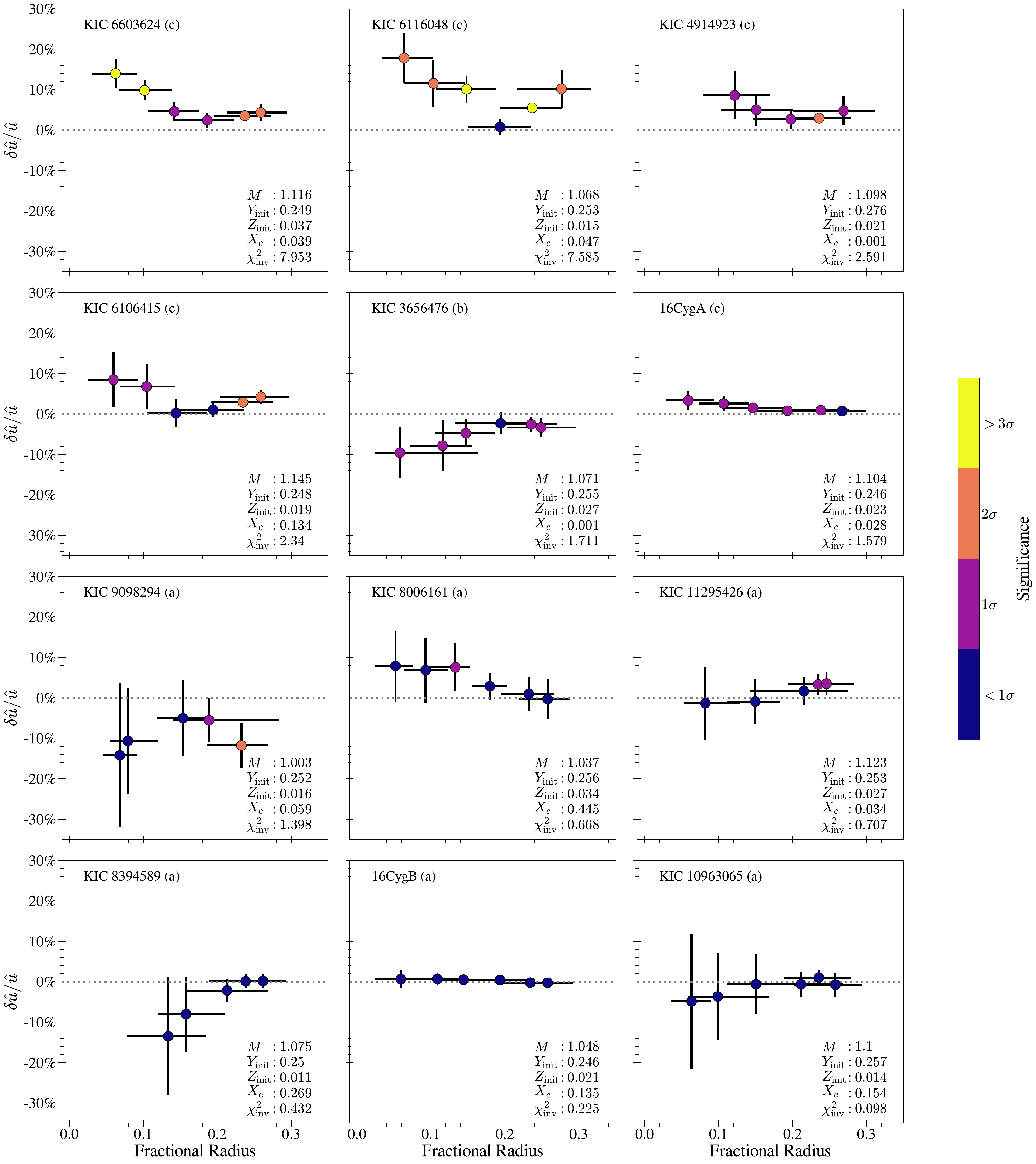}
    \caption{Comparisons of the internal structure of stars as revealed by asteroseismology and the structures of best-fitting stellar evolution models. Relative differences are given in terms of the dimensionless squared isothermal sound speed \(\hat u\) and span the near-core region of 0.05-0.3 away from the stellar center point. The points indicate the inferred value of \(\delta \hat{u}/\hat{u}\) between the star and the reference model at the target radius. The vertical error bars indicate the uncertainty of each inversion result from the propagation of the uncertainty of the observed frequencies. The horizontal error bars represent the full width at half maximum of the averaging kernel. The dashed horizontal line indicates complete agreement between the model and observations; points above this line imply that \(\hat{u}\) of the star is larger than that of the model. The color bar indicates the statistical significance of the inferred difference, with lighter colors showing more significant results. \rev{The letter after the star's identifier indicates which group the star is in, as described in the text.} The values given in the lower left of each plot indicate the mass (\(M)\), initial helium mass fraction \((Y_{\rm{init}})\), initial metallicity \((Z_{\rm{init}})\), and central hydrogen mass fraction \((X_{c})\) of each reference model. We also report the overall significance of the inversion results, \(\chi^{2}_{\rm{inv}}\).}
    \label{fig:all_inv_res}
\end{figure*}

Together, the \emph{Kepler} LEGACY and KAGES samples provide oscillation data for 95 stars. Since we are specifically searching for close matches to stars with radiative cores, we apply two criteria to our reference models: that they have a radiative core throughout their main-sequence evolution and that they have a \(\chi^2_{\rm{fit}} < 20\). We obtain suitable reference models for 34 stars. Of these, 12 have enough frequencies observed (approximately 35) to form well-localized averaging kernels.

Figure~\ref{fig:all_inv_res} shows the inversion results for each of these 12 stars as a function of the target radius. We define our relative differences such that a positive inversion result indicates a sound speed that is higher in the star than in the model. We provide more information about the reference model and averaging kernels of each star in Appendix~\ref{appendix: Inversion details}.

\rev{
The inversion results of the twelve stars in our sample can be broken down into three groups: (a) those where the \(\hat{u}\) of the best-fit model is in agreement with the observations, (b) those where the \(\hat{u}\) of the model is too high (resulting in an inversion result below zero), and (c) those where the \(\hat{u}\) is too low (resulting in an inversion result above zero). Taking into account the uncertainties of the inversion results, we identify six stars in group a, five stars in group b, and one star in group c. Thus, half of our sample show significant differences, which suggests that there are limitations in the physics of our reference models and that these limitations most often result in internal values of \(\hat{u}\) that are too low.}

\begin{figure*}
    \centering
    \plotone{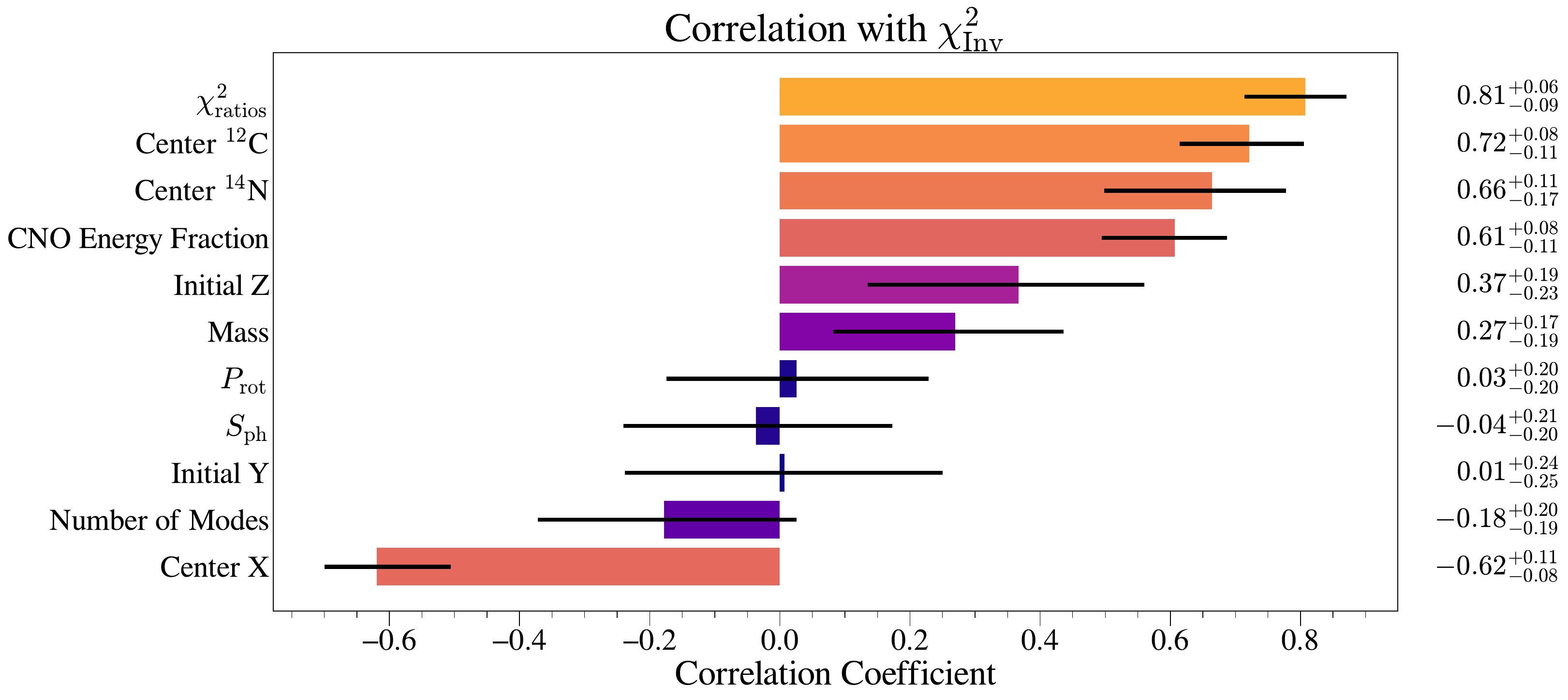}
    \caption{Spearman rank correlation between the maximum significance inversion result of each star and various properties of the reference model. The color correlates with the absolute value of the correlation coefficient, which is a measure of the strength of the correlation. The value of each correlation coefficient are provided on the right side of the figure. We estimate the uncertainty of each correlation coefficient using bootstrapping.}
    \label{fig:correlation} 
\end{figure*}

Now we seek to understand why some of our stellar models show good agreement in \(\hat{u}\) while others show significant disagreement. We search for correlations between \(\chi^2_{\rm{inv}}\) and properties of the reference model, as well as the surface rotation rate (\(P_{\rm{rot}}\)) and magnetic activity indicator (\(S_{\rm{ph}}\)) values for each star, as given by \citet{Santos_2018}. We calculate Spearman's rank correlation coefficient, \(\rho_{s}\), which describes the strength of the monotonic, but not necessarily linear,  correlation between two variables. 
These results are shown in Figure~\ref{fig:correlation}. We use bootstrapping to obtain estimates of the uncertainty of these coefficients.

The strongest correlation is with \rev{\(\chi^{2}_{\rm{ratios}}\) (\(\rho_{s} = 0.81\))}. This \(\chi^{2}\) is a measure of how well our reference model matches the observed frequency separation ratios \rev{($r_{10},r_{02}$)} of the observed star. These ratios are known to be insensitive to the surface effect, \citep{2003A&A...411..215R} and thus \(\chi^{2}_{\rm{ratios}}\)  serves as a different metric for how well the internal structure of a star is reproduced in the model. The strong correlation between \(\chi^{2}_{\rm{ratios}}\) and \(\chi^{2}_{\rm{inv}}\) reaffirms that the differences found from structure inversions are internal structure differences rather than problems with the near-surface layers.

We find significant positive correlations of the discrepancies between the star and stellar model with the central abundance of \(^{12}\)C \rev{(\(\rho_{s} = 0.72\))} and \rev{\(^{14}\)N (\(\rho_{s} = 0.66\))} of the model, as well as the amount of energy generated by the CNO cycle \rev{(\(\rho_{s} = 0.61\))}. A similar-strength correlation in the opposite direction is found with the central hydrogen abundance \rev{(\(\rho_{s} = -0.62\))}. That all of these properties have a similar strength of correlation is unsurprising, as they are mutually correlated.  When the central hydrogen value is lower, reactions other than the pp-chain can happen more easily. Primarily, this is an increase in energy generated by the CNO cycle. At the same time, a very small amount of energy is generated by the triple alpha process. This is not significant compared to the total energy generation of the star, but it does increase the equilibrium abundance of \(^{12}\)C. \rev{Additionally, the CNO-II pathway converts \(^{16}\)O into \(^{14}\)N which increases the equilibrium abundance of \(^{14}\)N. } In general, we see a moderate correlation between the significance of the \(\hat{u}\) differences inferred by inversions and more evolved main sequence stars. 

\rev{
\subsection{Individual Stars}}
\rev{We now discuss the results of both our forward modeling and inversion procedures for a few individual stars.} 
\subsubsection{\rev{16~Cyg~A and 16~Cyg~B}}
\begin{figure*}
    \centering
    \gridline{\fig{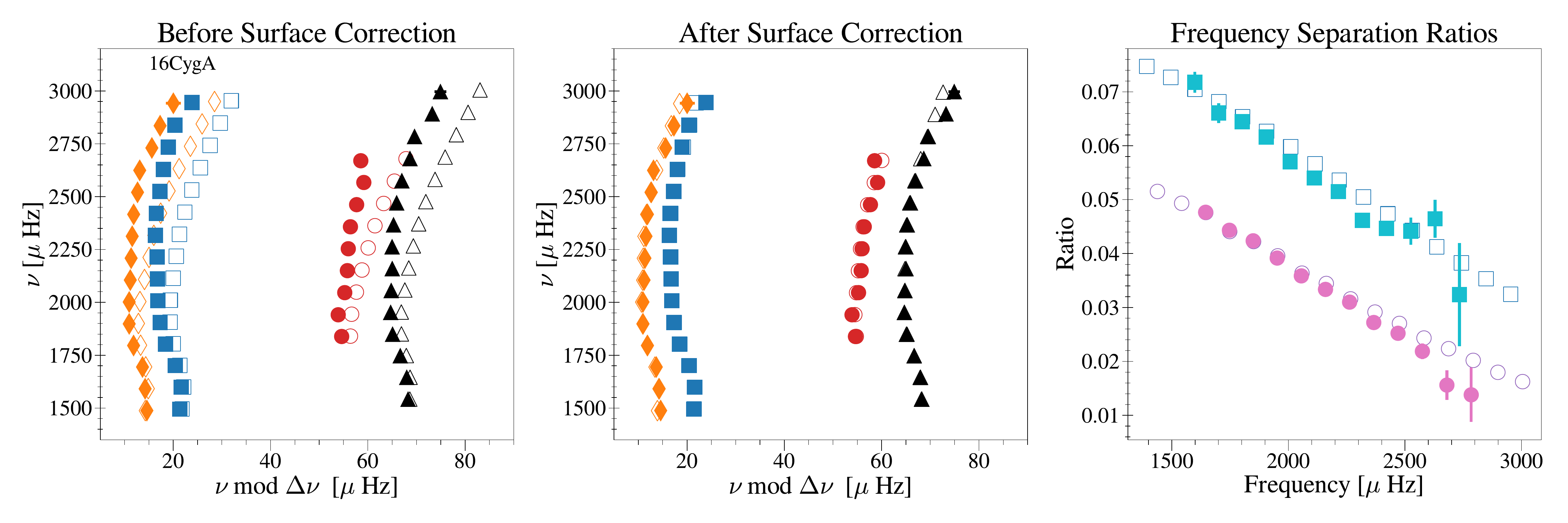}{\textwidth}{}}
    \gridline{\fig{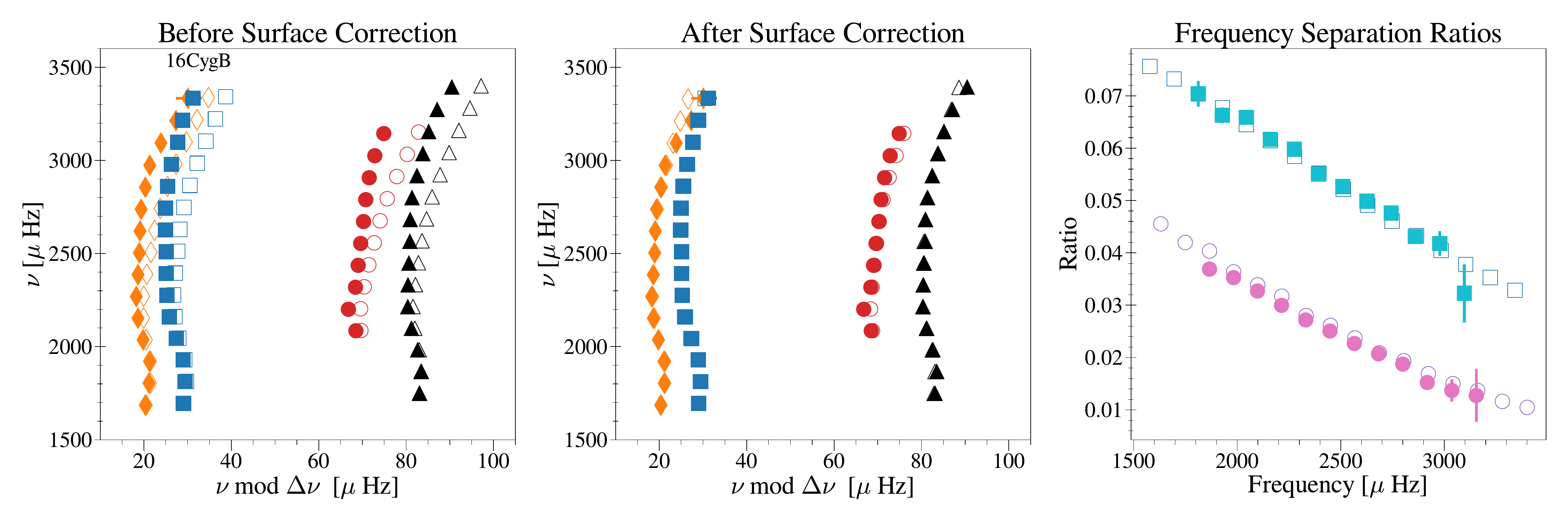}{\textwidth}{}}
    \caption{\rev{Modeling results for 16~Cyg~A (top) and 16~Cyg~B (bottom). In each case, the left plot shows the Frequency \'Echelle diagram comparing the frequencies of the reference model to the observations before applying any correction to account for surface effects. The center panel compares the reference model frequencies after applying the two-term surface correction from \citet{2014A&A...568A.123B}. The color and shape indicate the spherical degree \(l\): 
    0 (blue squares), 1 (black triangles), 2 (orange diamonds), and 3 (red circles). The right plot shows the frequency separation ratios \(r_{10}\) 
    (pink) and \(r_{02}\) (light blue). In all plots, the open points represent the values from the reference model and the filled points represent the observed values.} }
    \label{fig:16Cyg_freqs} 
\end{figure*}

\begin{figure*}
    \epsscale{0.8} 
    \centering
    \plottwo{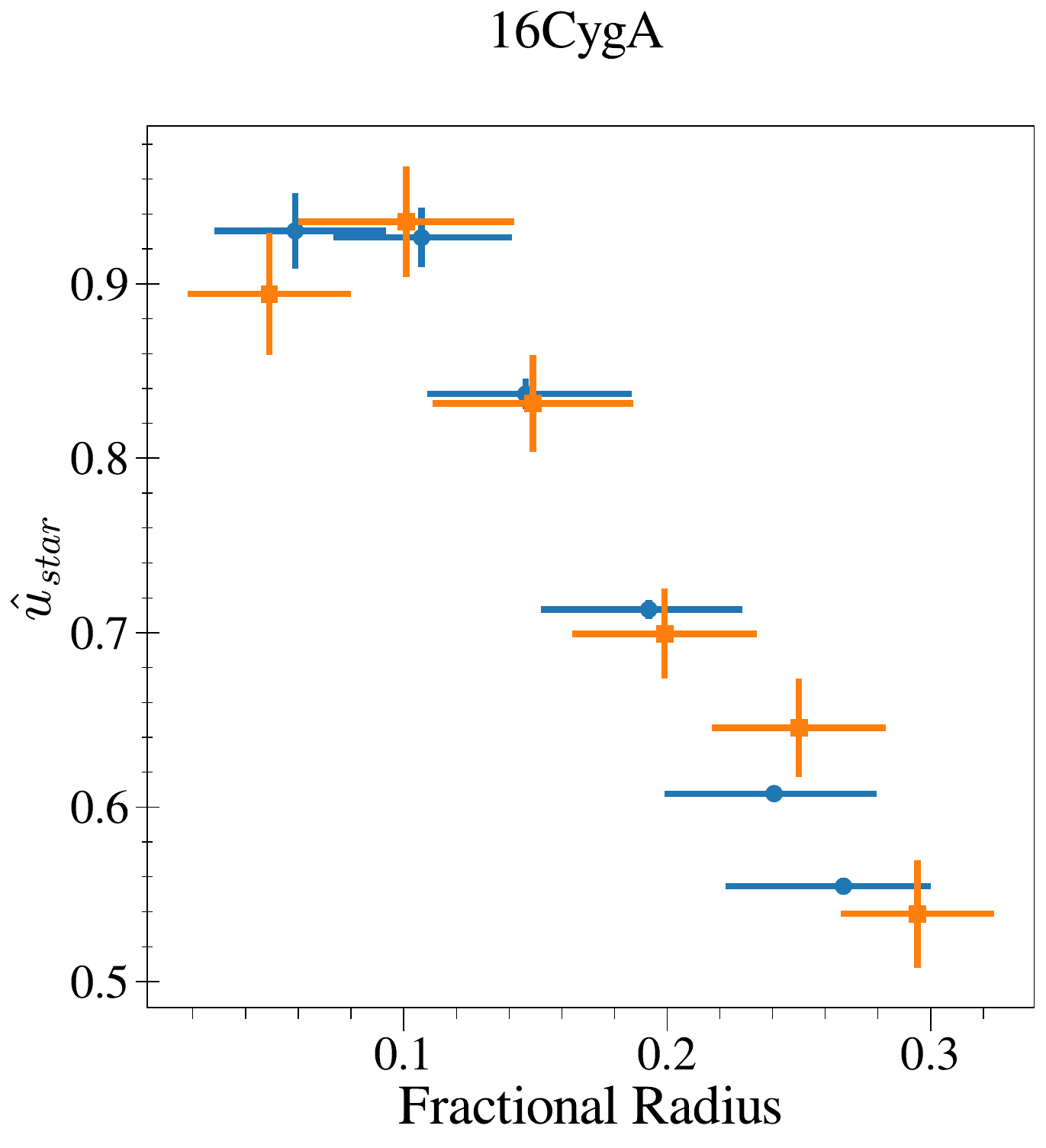}{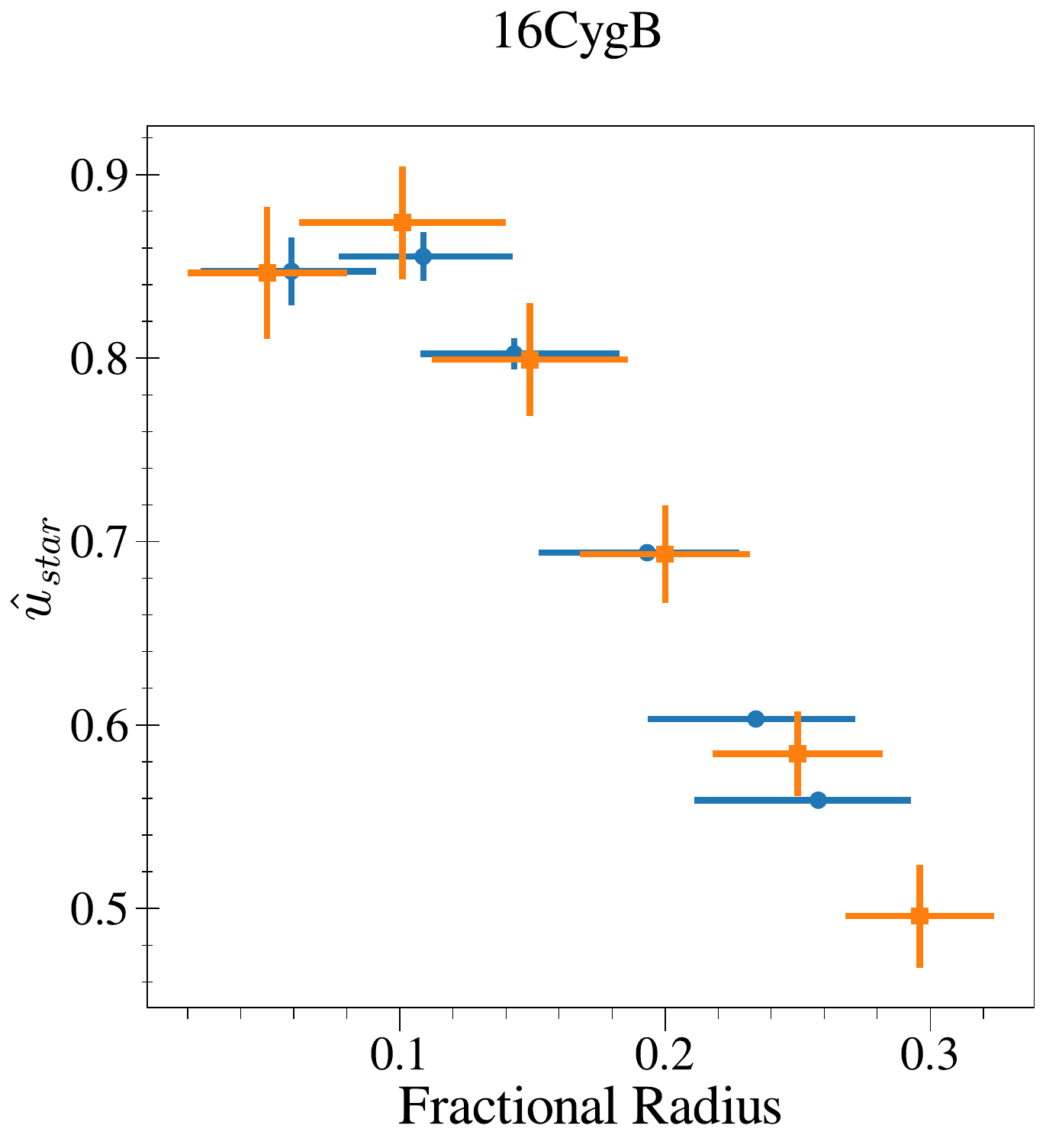}
    \caption{\rev{Inversion results for 16~Cyg~A (left) and 16~Cyg~B (right). The blue points show the inversion results from this work. The orange points are the results from \citet{2017ApJ...851...80B}. Since they report \(u\) we use their reported values of M and R to calculate \(\hat{u}\).} }
    \label{fig:16Cyg}
\end{figure*}

\rev{
First, we focus on the solar analogs 16~Cyg~A and 16~Cyg~B. In Figure~\ref{fig:16Cyg_freqs} we compare the frequencies (before and after surface-term corrections) and frequency separation ratios of our reference models with the observations. As these are two of the most well-studied main sequence stars in the \emph{Kepler} field, they have already been studied using structure inversions by  \citet{2017ApJ...851...80B,2022A&A...661A.143B}. In our results, as well as the two previous studies, there is excellent agreement between the models and the observations. In the case of 16~Cyg~B, the models used in all three works are within 1\(\sigma\) agreement with observations. For the case of 16~Cyg~A, our inversions show differences that are less than 1.5\(\sigma\), which is similar to the values obtained by \citet{2017ApJ...851...80B} and \citet{2022A&A...661A.143B}. 
\citet{2017ApJ...851...80B} report their inferred \(u\) values, as well as their inferred values of the stellar mass and radius, which allows us to compare \(\hat{u}\) values directly, as shown in Figure~\ref{fig:16Cyg}. All the points for 16~Cyg~B are in good agreement. For 16~Cyg~A, there is slight disagreement at a target radius of 0.25, but it is not significant. Despite the use of different reference models, a different implementation of OLA, and different inversion parameters, we agree on the internal sound speed profiles of both 16~CygA and 16~Cyg~B.}

\rev{
\subsubsection{KIC 6116048 and KIC 6603624}} \label{sec:611_660}

\begin{figure*}
    \centering
    \gridline{\fig{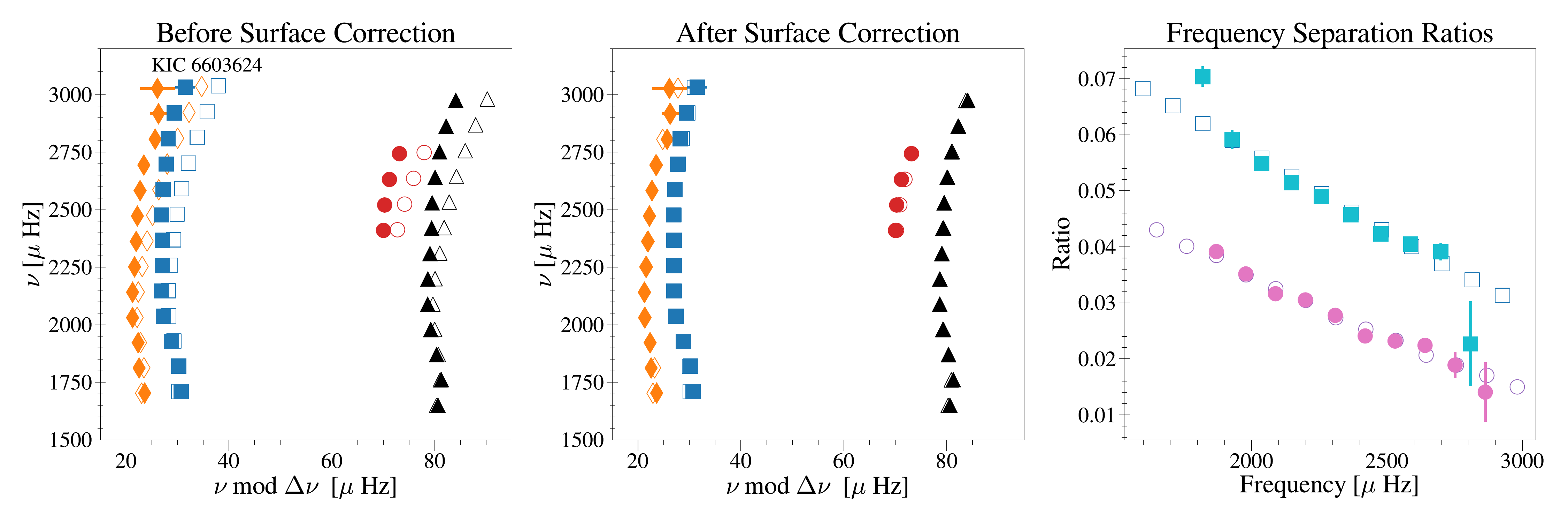}{\textwidth}{}}
    \gridline{\fig{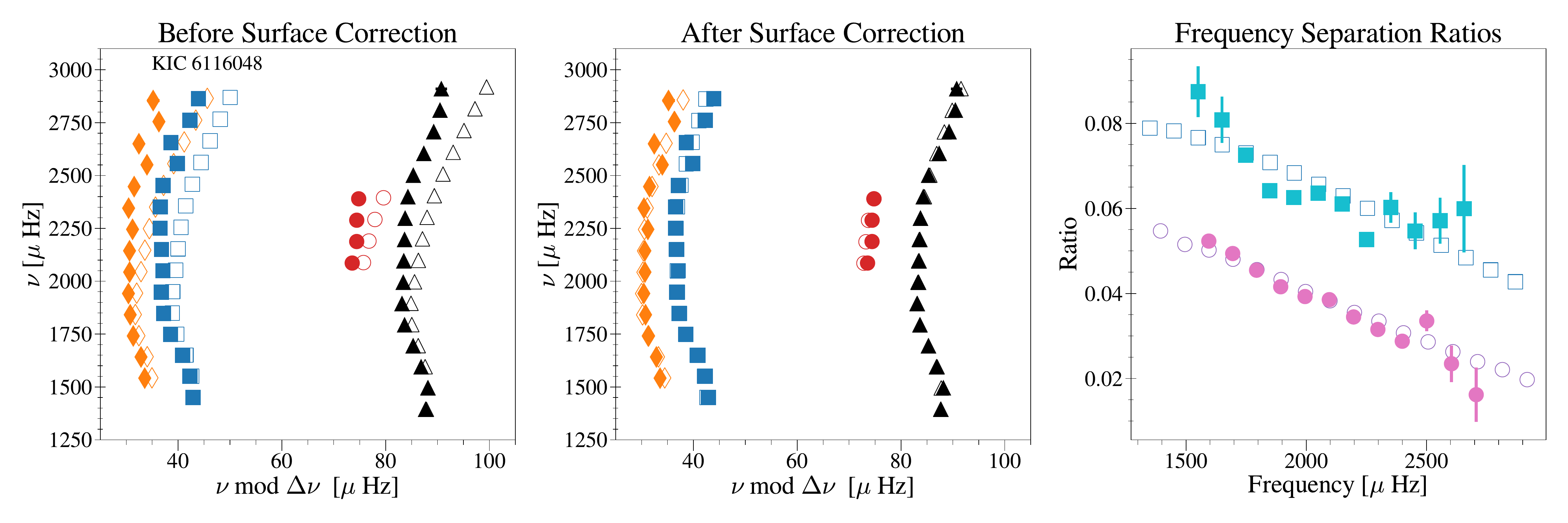}{\textwidth}{}}
    \caption{Modeling results for KIC~6603624 (top) and KIC~6116048 (bottom). All symbols have the same meaning as in Figure~\ref{fig:16Cyg_freqs}.}  
    \label{fig:611-660_freqs} 
\end{figure*} 

\begin{figure*} 
    \centering 
    \plottwo{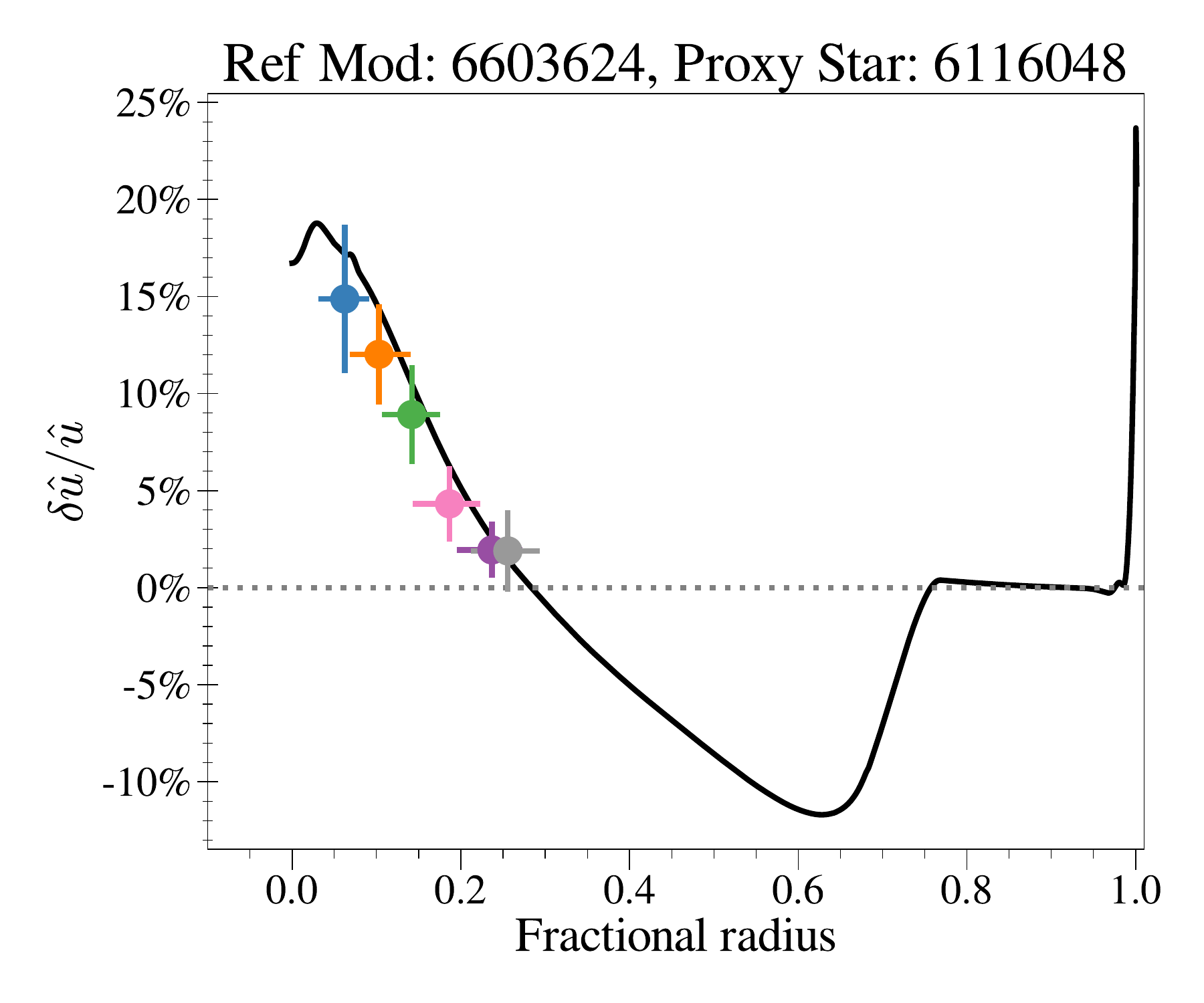}{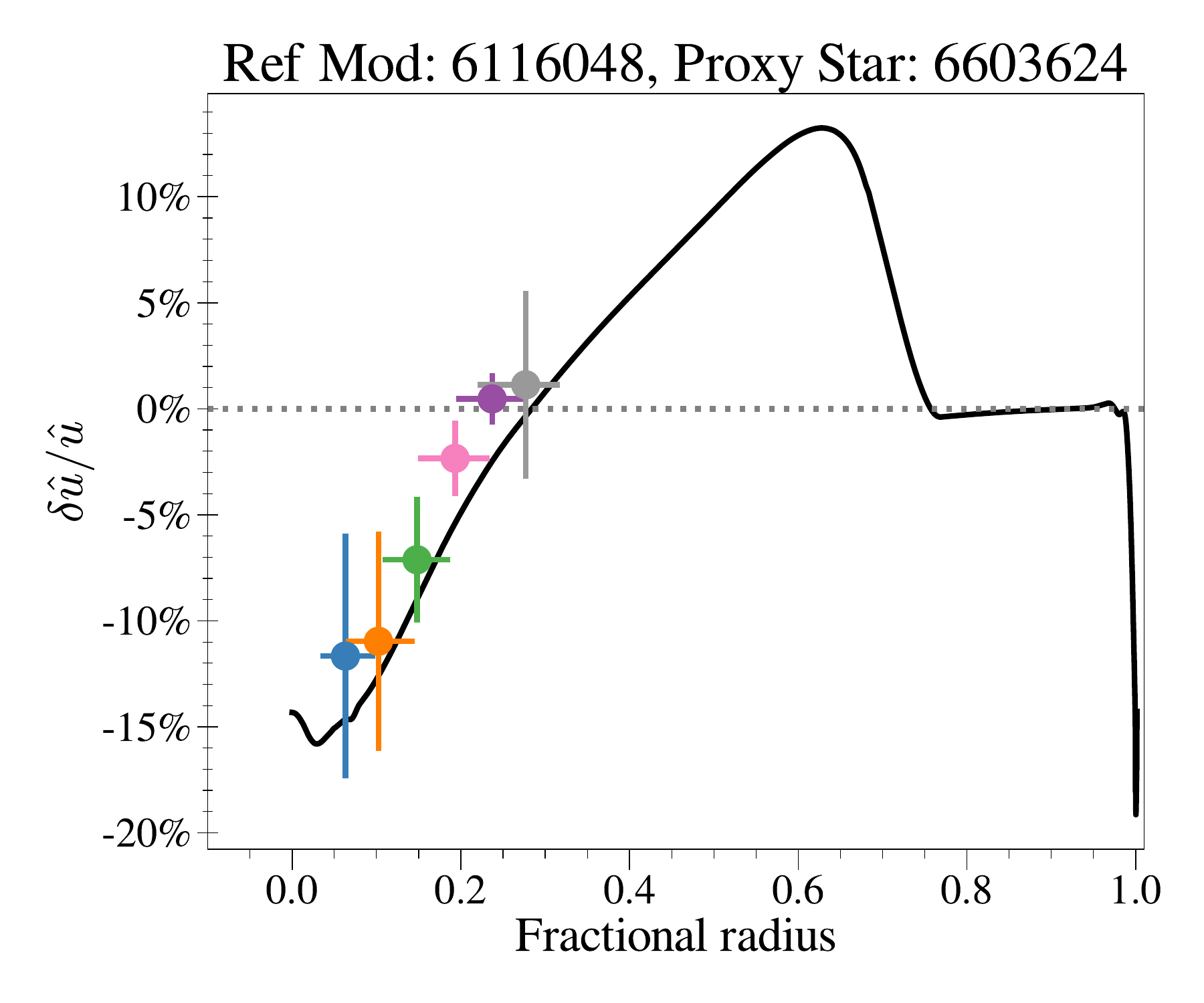}
    \caption{\rev{Model-model inversions to recover the \(\hat{u}\) difference between the model for KIC~6603624 and KIC~6116048. The left figure shows the result of using KIC~6603624 as the reference model, and the right figure shows the result of using KIC~6116048 as the reference model. In both plots, the black line represents the true value of \(\delta \hat{u}/\hat{u}\) and the colored points show the result of the inversion. Different target radii are shown in different colors and correspond to the color of the averaging and cross term kernels shown in Figures~\ref{fig:avg_kern_all} and \ref{fig:cross_kern_all}.}  }
    \label{fig:611_660_inv} 
\end{figure*}

\rev{
We now turn to the two stars in our sample that show the largest differences with respect to our models: KIC~6603624 and KIC~6116048. We show the frequencies and frequency separation ratios in Figure~\ref{fig:611-660_freqs}. Both of these stars have points where our inversions infer internal sound speed differences greater than 10 percent, and in contrast to other stars in the sample, these large differences are significant compared to their uncertainties, so first we verify that our inversions are able to recover differences of this magnitude. The \(\delta \hat{u}/\hat{u}\) between our reference model for KIC~6116048 and our reference model for KIC~6603624 reach \(\sim\)15\% in the region probed by structure inversions, and so we test our averaging kernels by attempting to recover the difference between the two models. We do this twice, once with KIC~6603624 as the reference model and then again using KIC~6116048 as the reference model. The result of these inversions are shown in Figure~\ref{fig:611_660_inv}. Both sets of averaging kernels infer the correct shape of the true \(\delta \hat{u}/\hat{u}\) curve. The averaging kernels of KIC~6603624 infer the correct value of \(\delta \hat{u} /\hat{u}\) within the uncertainties at every target radius. This is not the case for the averaging kernels for KIC~6116048, where two points differ from the correct value by \(\sim 2 \sigma\). Nevertheless, we conclude that our inversion procedure is able to recover differences around 15\%.}

\subsection{Exploring the effects of microphysics}
\begin{figure*} 
    \epsscale{1.2}
    \plotone{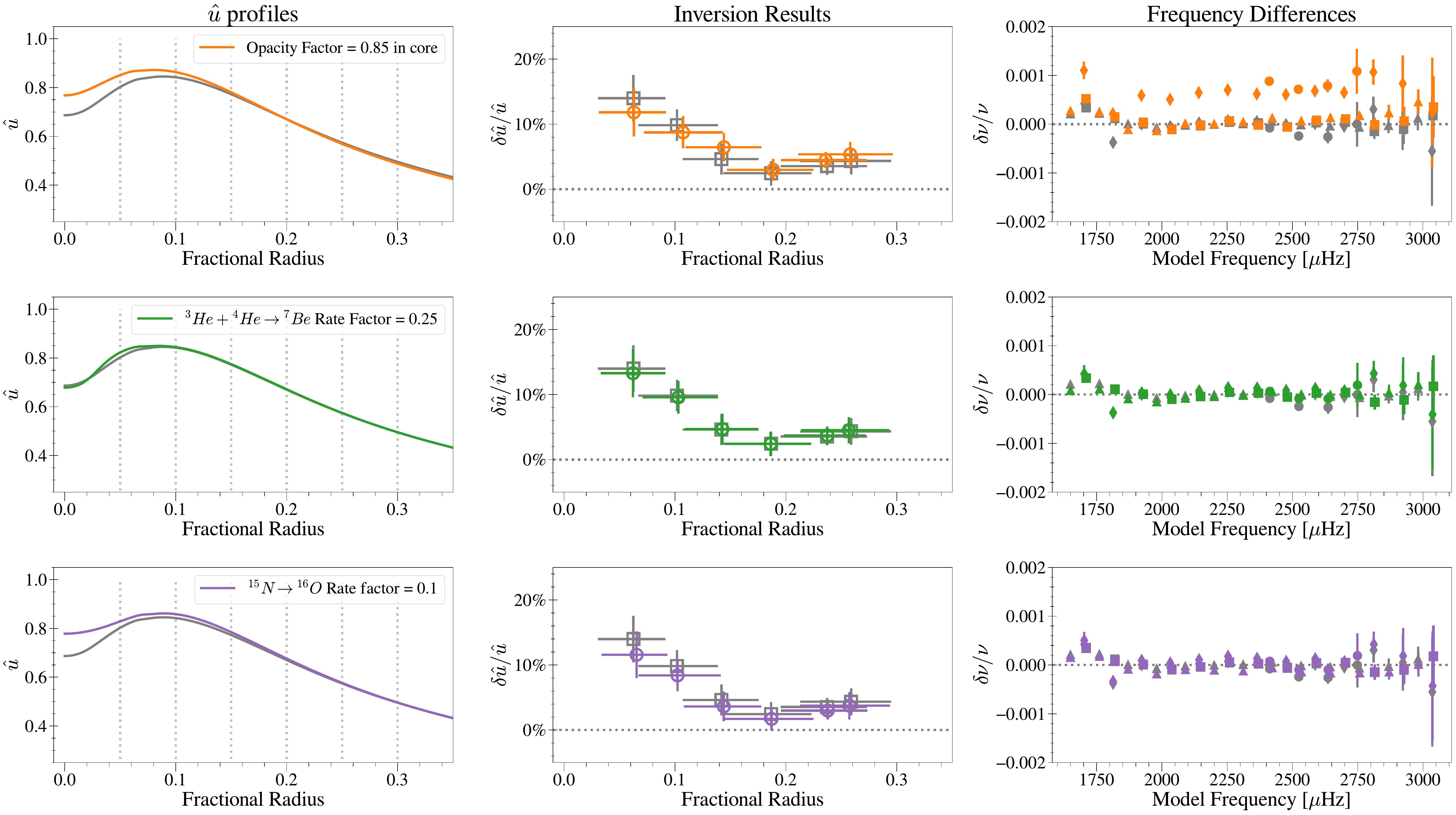}
    \caption{\rev{Results of modifying the physics used to evolve each model.  The \(\hat{u}\) profile of each model is shown in the left plot, with light gray dashed vertical lines to indicate the target radii of the inversions. The center plot shows the result of structure inversions using each model as the reference model. The right plot shows the frequency differences between each model and the observed modes of KIC~6603624. The shape of the marker denotes the spherical degree of the mode, with $l=0,1,2,3$ denoted by squares, triangles, diamonds, and circles respectively. In each plot, gray lines and points represent the values of the original reference model of KIC~6603624. }}
    \label{fig:660_tests} 
\end{figure*}

\begin{figure*} 
    \epsscale{1.2}
    \plotone{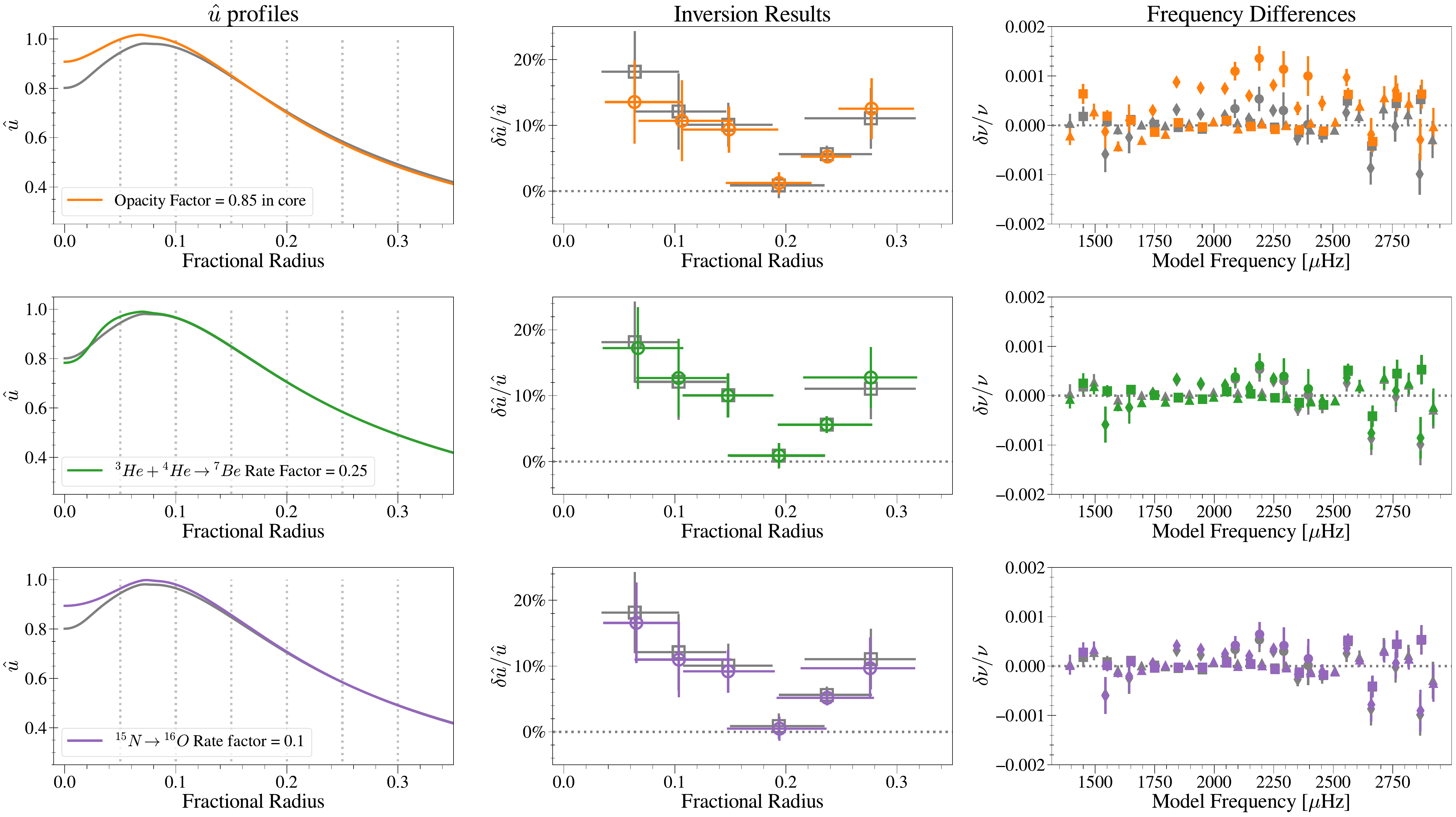}
    \caption{\rev{Results of modifying the physics used to evolve each model of KIC~6116048. All colors and symbols have the same meaning as in Figure~\ref{fig:660_tests}.}}
    \label{fig:611_tests} 
\end{figure*}

\rev{We now explore several changes to the microphysics in our models in an attempt to reduce the sound speed differences inferred by our inversions. A full investigation of the microphysics across all twelve of the stars studied here is beyond the scope of this work, and hence we focus on KIC~6603624 and KIC~6116048, the two stars discussed in section \ref{sec:611_660}. For each star, }
we create three new models using the same mass, initial composition, and mixing-length parameter as our original reference model, although we allow these new models to have a different central hydrogen abundance. For the first model, motivated by the correlation to CNO energy production, we multiply the rate of the \({}^{14}\rm{N} + p \rightarrow {}^{15}\rm{O} + \gamma\) reaction by a factor of 0.1. For the second model, we multiply the ppII/ppIII rate \(^3\rm{He} + {}^4\rm{He} \rightarrow {}^7\rm{Be} + \gamma\) by a factor of 0.25. For the last model, we modify the opacity by a factor of 0.85 in the parts of the model with \(\log T > 6.7\). For each of these three new tracks, we select a new reference model using the fitting procedure discussed in Section~\ref{sec:modeling}. 
\rev{
Figures~\ref{fig:660_tests} and \ref{fig:611_tests} show the results of these changes for KIC~6603624 and KIC~6116048, respectively. For each change in the microphysics we show the \(\hat{u}\) profiles of each new reference model as well as the result of structure inversions.} 
The changes to the core opacity and the \({}^{14}\rm{N} + p \rightarrow {}^{15}\rm{O} + \gamma\) reaction rate both increase \(\hat{u}\) from the original reference model, with the opacity change resulting in a larger difference both to the central \(\hat{u}\) and the frequency differences computed with respect to the observations. The change caused by modifying the \({}^3\rm{He} + {}^4\rm{He} \rightarrow {}^7\rm{Be} + \gamma\) reaction rate results in a smaller change to the \(\hat{u}\) that is only apparent inside \rev{$r/R < 0.07$}.  
As expected, when we apply changes that increase the internal \(\hat{u}\), the \(\hat{u}\) difference inferred by the structure inversion decreases. We find better agreement in the \(\hat{u}\) profile even when the fit of the model is worse than our original reference model. These changes improve our models at the deepest target radii, but have little effect at the larger radii probed by inversions.

\section{Conclusions} \label{sec:concul}
Here, we have used asteroseismology to infer the detailed core structure of the best solar-type stars observed by the \textit{Kepler} mission. We focused on main sequence-stars with radiative cores and expanded the number of such stars studied with structure inversions from 2 to 12. 
After obtaining our reference models from a grid created using MESA, we use a set of calibration models to obtain our inversion parameters. We then use these inversion parameters to infer the relative difference in dimensionless squared isothermal sound speed between our reference model and the target star. \rev{In our sample, we identify three groups: those where the \(\hat{u}\) of our reference model agrees with the observed star  (group a, 6 stars), those where the \(\hat{u}\) of our model is higher than that of the star (group b, 1 star), and those where the \(\hat{u}\) of our model is lower than that of the star (group c, 5 stars).} We also find significant correlations in our results, suggesting that our models of older main-sequence stars with more energy being generated by the CNO cycle have larger differences between model and star. To explore how changing the microphysics affects our inversion results, we tested the effects of changing nuclear reaction rates and core opacities, \revv{for the two stars} with the most significant differences. These changes to the microphysics reduced the discrepancy between model and star at the innermost target radii.

In future work, we aim to extend our analysis to an even broader set of stars, including main-sequence stars with convective cores and more evolved stars with mixed-mode oscillations. Main-sequence stars with convective cores are particularly interesting since their dominant source of energy is the CNO cycle. Thus, they are  a natural next step to explore the correlations found in this work between the inferred sound speed differences and CNO energy production.

\begin{acknowledgments}
\rev{We thank the anonymous referee for their constructive feedback that improved the manuscript considerably.}
The research leading to the presented results has received funding from the ERC Consolidator Grant DipolarSound (grant agreement \#101000296). \rev{SB acknowledges NSF grant AST-2205026. SB would also like to thank the Heidelberg Institute of Theoretical Studies for their hospitality during the early days of this project.}
This paper includes data collected by the Kepler mission. Funding for the Kepler mission is provided by the NASA Science Mission Directorate. 
In addition, this work has made use of data from the European Space Agency (ESA) mission {\it Gaia} (\url{https://www.cosmos.esa.int/gaia}), processed by the {\it Gaia}
Data Processing and Analysis Consortium (DPAC, \url{https://www.cosmos.esa.int/web/gaia/dpac/consortium}). Funding for the DPAC has been provided by national institutions, in particular the institutions
participating in the {\it Gaia} Multilateral Agreement.
We have also used the gaia-kepler.fun crossmatch database created by Megan Bedell.
\end{acknowledgments}


\facilities{Kepler, Gaia}

\clearpage
\appendix
\twocolumngrid
\section{Reference Model Parameters} \label{appendix:ref_models} 
\rev{Table~\ref{tab:obs} provides the non-seismic constraints that were used to find the reference models of the 12 main sequence stars with radiative cores discussed in this work.} Table~\ref{tab:RefModInfo} provides the model parameters for each reference model. 

\begin{deluxetable}{lccc} 
    \tablecaption{Non-seismic observations} 
    \label{tab:obs} 
    \tablehead{\colhead{Star} & \colhead{$T_{\rm{eff}}$ [K]} & \colhead{[Fe/H]} & \colhead{Luminosity [$\rm{L}_\odot$]}} 
    \startdata
KIC 6603624 & 5602$\pm$100 & 0.29$\pm$0.1 & 1.241$\pm$0.018\\ 
KIC 6116048 & 6012$\pm$100 & -0.26$\pm$0.1 & 1.862$\pm$0.006\\ 
KIC 4914923 & 5823$\pm$100 & 0.12$\pm$0.1 & 2.135$\pm$0.035\\ 
KIC 6106415 & 5975$\pm$100 & -0.09$\pm$0.1 & 1.882$\pm$0.006\\ 
KIC 3656476 & 5664$\pm$100 & 0.28$\pm$0.1 & 1.719$\pm$0.028\\ 
16CygA & 5777$\pm$100 & 0.01$\pm$0.1 & 1.563$\pm$0.005\\ 
KIC 9098294 & 5869$\pm$100 & -0.18$\pm$0.1 & 1.413$\pm$0.007\\ 
KIC 8006161 & 5422$\pm$100 & 0.32$\pm$0.1 & 0.646$\pm$0.005\\ 
KIC 11295426 & 5784$\pm$100 & 0.04$\pm$0.1 & 1.62$\pm$0.01\\ 
KIC 8394589 & 6051$\pm$100 & -0.4$\pm$0.1 & 1.853$\pm$0.007\\ 
16CygB & 5734$\pm$100 & -0.01$\pm$0.1 & 1.221$\pm$0.005\\ 
KIC 10963065 & 6100$\pm$100 & -0.22$\pm$0.1 & 1.934$\pm$0.007\\ 
    \enddata
\end{deluxetable}

\begin{deluxetable}{ccccccc}
\tablecaption{Reference Model Parameters} 
\label{tab:RefModInfo} 
\tablehead{\colhead{Star} & \colhead{$M [\rm{M}_{\odot}]$} & \colhead{$Y_{\rm{initial}}$} & \colhead{$Z_{\rm{initial}}$} & \colhead{$\alpha_{\rm{mlt}}$} & \colhead{$X_{c}$} & \colhead{$\chi^{2}_{\rm{fit}}$}}

\startdata
   KIC 6603624 & 1.116 &     0.249 &     0.037 &             2.111 &   0.039 &               4.751 \\
   KIC 6116048 & 1.068 &     0.253 &     0.015 &             2.227 &   0.047 &               4.238 \\
   KIC 4914923 & 1.098 &     0.276 &     0.021 &             1.849 &   0.001 &               5.441 \\
   KIC 6106415 & 1.145 &     0.248 &     0.019 &             2.341 &   0.134 &               3.113 \\
   KIC 3656476 & 1.071 &     0.255 &     0.027 &             1.755 &   0.001 &              14.846 \\
    16CygA & 1.104 &     0.246 &     0.023 &             2.145 &   0.028 &               4.661 \\
    KIC 9098294 & 1.003 &     0.252 &     0.016 &             2.173 &   0.059 &               4.012 \\
  KIC 11295426 & 1.123 &     0.253 &     0.027 &             1.967 &   0.034 &               3.387 \\
   KIC 8006161 & 1.037 &     0.256 &     0.034 &             2.265 &   0.445 &               2.623 \\
   KIC 8394589 & 1.075 &     0.250 &     0.011 &             2.266 &   0.269 &               3.541 \\
    16CygB & 1.048 &     0.246 &     0.021 &             2.281 &   0.135 &               3.410 \\
  10963065 & 1.100 &     0.257 &     0.014 &             2.278 &   0.154 &               1.851 \\\enddata
\end{deluxetable}

\section{Inversion Details}
\revv{Here we provide details on how we chose our inversion parameters and how we calculate the non-dimensional frequency differences used in our inversions. Additionally, we present the results of applying our modeling and inversion methods to degraded solar data. }

\revv{\subsection{Inversion Parameter Selection} \label{appendix:mu} }
\revv{For each target radius we find the value of \(\mu\) that minimizes:}
\rev{
\begin{equation} 
\mathcal{M} =  \left< \left[\left(\frac{\delta \hat{u}}{\hat{u}}\right)_{\rm{inv}} - \left(\frac{\delta \hat{u}}{\hat{u}}\right)_{\rm{True}} \, \right] + \sigma_{\rm{inv}} \right>_{\rm{set}}  
\label{equ:mu_metric} 
\end{equation}}
\revv{
where the angle brackets denote a mean across the set of calibration models,  \(\left(\delta \hat{u}/\hat{u}\right)_{\rm{inv}}\) is the sound speed difference inferred by the inversion, \(\left(\delta \hat{u}/\hat{u}\right)_{\rm{True}}\) is the true sound speed difference, and the uncertainty of the inversion result is \(\sigma^2_{\rm{inv}} = \sum_i c_i^2 \sigma_{i}^2\), where \(\sigma_i\) is the relative uncertainty of the \(i\)th mode. This is not the uncertainty reported in our final results, as it does not account for uncertainty correlation introduced by our surface term and mean density corrections (see Section~\ref{sect:MR}). In general, the term in the square brackets dominates, as with increasing values of \(\mu\) the quality of the averaging kernel degrades faster than the uncertainty of the final result is reduced.  We minimize \(\mathcal{M}\) separately for each target radius, and so the value of \(\mu\) can vary between different target radii of the same star.  This optimization is more stable when only one variable is minimized, and so we set the cross-term trade-off parameter \(\beta=0\). }

\begin{figure*} 
\centering 
\plotone{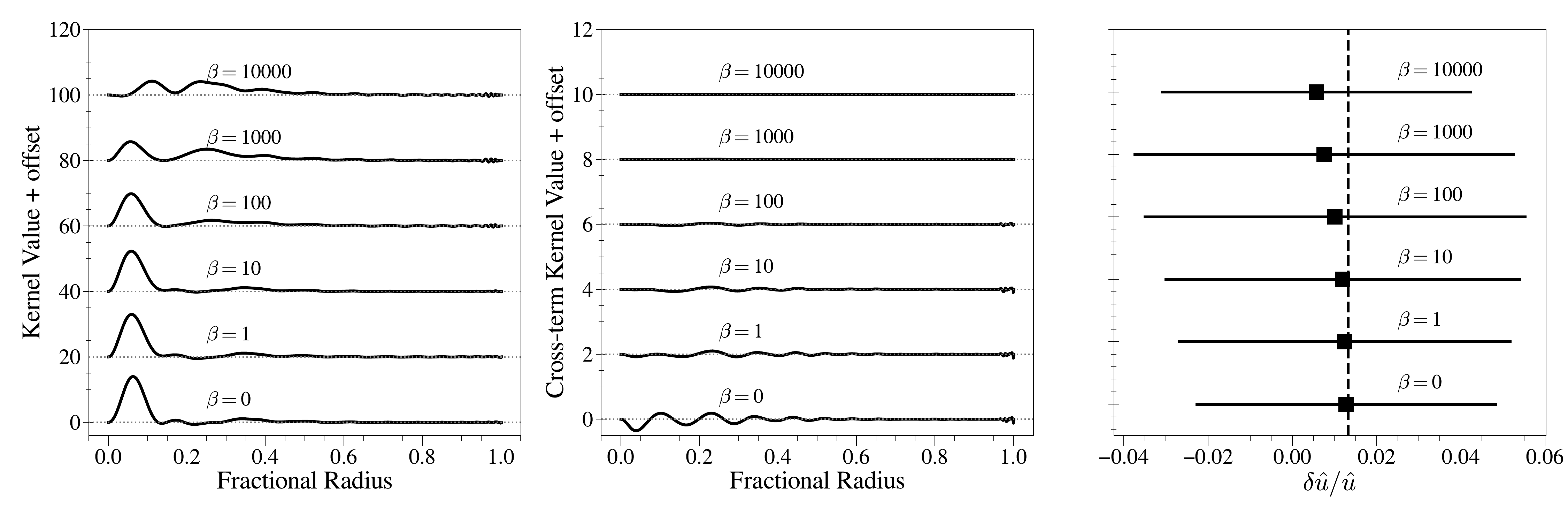} 
\caption{\rev{Results of varying \(\beta\) for the innermost target radius of KIC~6603624. The left (center) panel shows the averaging kernels (cross-term kernels)  that result from the indicated value of \(\beta\). The right panel shows the result of a representative model-model inversion for each value of \(\beta\). The true value of \(\delta \hat{u}/\hat{u}\) is indicated by the vertical line. The error bars show the uncertainty on the inversion result.} }
\label{fig:beta_vary}
\end{figure*} 

\revv{As the effect of this choice is similar for all target radii across all stars in our sample, we use the innermost target radius of KIC~6603624 as an example. }Figure~\ref{fig:beta_vary} shows the averaging kernels, cross-term kernels, and model-model inversion results for several values of \(\beta\) using the same value of \(\mu\). Increasing \(\beta\) has the expected effect of damping the cross-term kernel, however it also reduces the quality of the averaging kernel, making it less localized. This is particularly noticeable when \(\beta = 10\,000\). As the model-model inversions show, the results are much more sensitive to the quality of the averaging kernel than to the amplitude of the cross-term kernel, and so we conclude that setting \(\beta=0\) is justified.

\subsection{Mean Density Scaling} \label{appendix:mean density} 
To mitigate the effect of a difference in mean density between a star and its model, we calculate the dimensionless frequency differences before applying our structure inversions. 
One method of obtaining this difference was proposed in \citet{2003Ap&SS.284..153B} and used by \citet{2021ApJ...915..100B}. This approach notes that the proportional scaling with mean density shows up as a constant offset in the frequency differences. This constant offset can be approximated by taking a weighted mean of the frequency differences. This term can then be subtracted from the raw frequency differences to remove any differences due to mean density. We have found that this approximation is valid only when the frequency differences due to different mean densities are larger than the differences resulting from structure differences. 
Thus, in this work, we take a different approach and use the large frequency separation of the star and its reference model to calculate a dimensionless frequency difference.
The dimensionless frequency is 
\begin{equation}
\hat{\nu} = \sqrt{\frac{R^3}{GM}}\; \nu
\end{equation}
where \(R\) is the stellar radius, \(M\) is the stellar mass, and \(G\) is the gravitational constant. For two stars with stellar radii \(R_1, R_2\) and stellar masses \(M_1, M_2\) the dimensionless relative frequency difference of any given mode is 
\begin{equation} 
\label{equ:non-dimdif-long}
\frac{\delta \hat{\nu}}{\hat{\nu}} = \frac{\sqrt{\frac{R_1^3}{GM_1}} \nu_1 - \sqrt{\frac{R_2^3}{GM_2}} \nu_2}{\sqrt{\frac{R_2^3}{GM_2}} \nu_2} = \sqrt{\frac{R_1^3}{M_1} \frac{M_2}{R_2^3}} \frac{\nu_1}{\nu_2} - 1
\end{equation} 
Since \(\Delta \nu \propto \sqrt{M/R^3}\), 
\begin{equation} 
\frac{\Delta \nu_2} {\Delta \nu_1} \approx \sqrt{\frac{M_2}{R_2^3} \frac{R_1^3}{M_1}}, 
\end{equation}
with this Equation~\ref{equ:non-dimdif-long} reduces to 
\begin{equation} 
\frac{\delta \hat{\nu}}{\hat{\nu}} \approx \frac{\Delta \nu_2}{\Delta \nu_1} \frac{\nu_1}{\nu_2} - 1.
\end{equation}

\begin{figure} 
\plotone{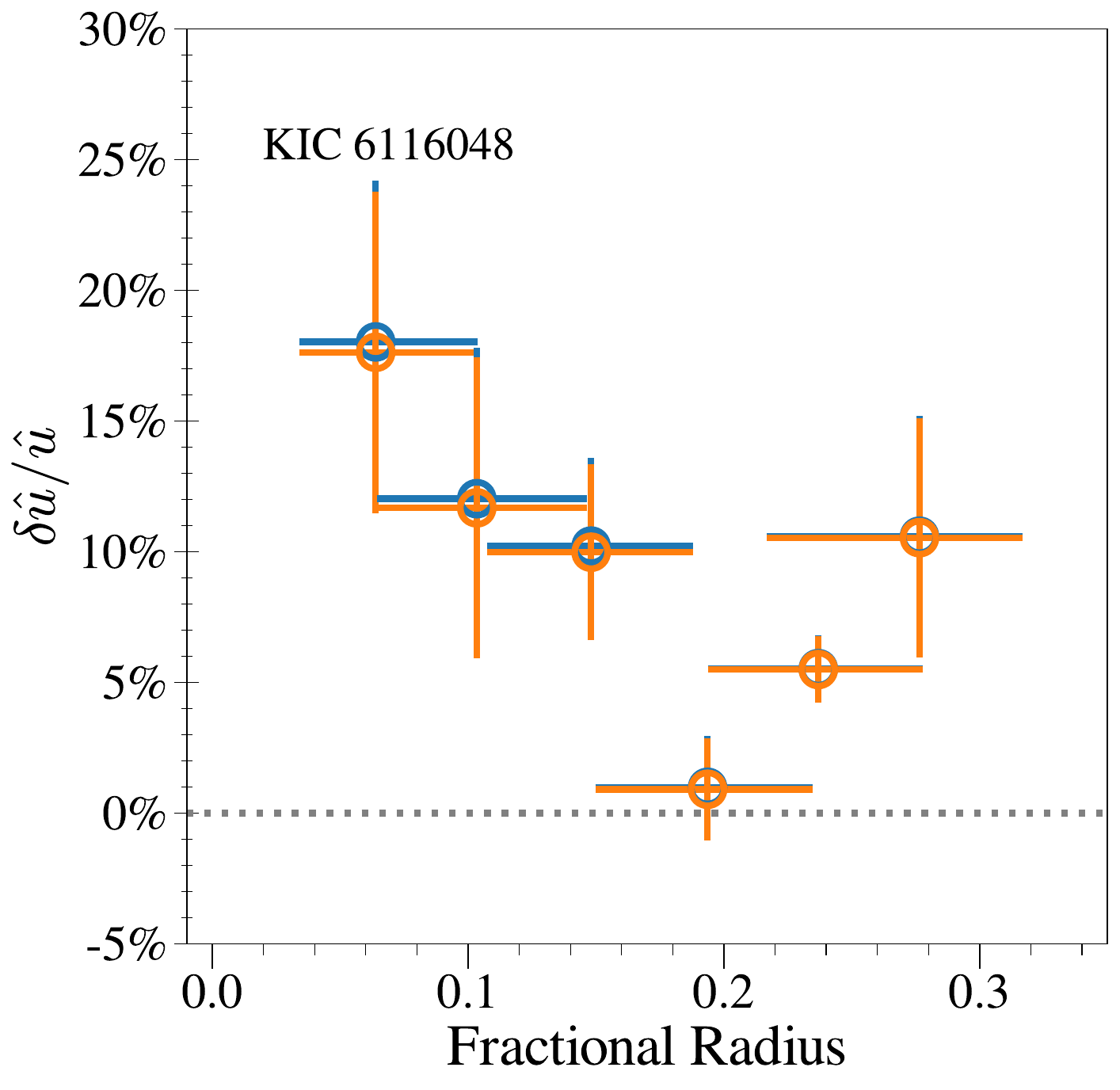} 
\caption{Inversion results for KIC~6116048 found using two different methods of calculating the dimensionless frequency differences. The blue points show the results when frequencies are scaled using the large frequency separations, described in Section \ref{sec:inversions}, and the orange points use the differences calculated from a weighted mean, described in \cite{2003Ap&SS.284..153B}. }
\label{fig:non-dim_comp} 
\end{figure}

While this method results in different values for the dimensionless frequency differences, the effect on the inversion result is small compared to the uncertainty of the inversion result, as seen for KIC~6116048 in Figure~\ref{fig:non-dim_comp}. This small difference can be understood by carrying forward the effect of a small difference in mean density though the inversion procedure. 

A difference of mean density shows up as a constant offset when calculating the dimensionless frequency differences. Mathematically, this is expressed as  
\begin{equation} 
\frac{\delta \hat{\nu}_i}{\hat{\nu}_i} = \frac{\delta \nu_i}{\nu_i} + \frac{\delta q}{q},
\end{equation} 
where \(\delta q/q\) is the offset introduced by a difference in mean density. 
As \(\delta q/q\) is independent of the frequencies, its contribution to the final inversion result will be 
\begin{equation} 
\left(\sum c_i \right) \frac{\delta q}{q}.
\label{equ:dim_err} 
\end{equation} 
Thus, the error introduced by a mismatch in mean density is proportional to the sum of the inversion coefficients for each target radii. We do not explicitly try to minimize this sum; however, the uncertainty of each result, which we do attempt to minimize, depends on the magnitude of the inversion coefficients. Thus, in the process of a standard inversion, we reduce the effect of a difference in mean density. 

Equation~\ref{equ:dim_err} also suggests a check to determine if a difference in mean density is the dominant difference present in our inversion results. While the sum of the coefficients will be different for each target radius, \(\delta q/q\) will be the same. Thus, if the mean density differences are dominating the inversion results, a plot of the inversion results at each target radii divided by the sum of the coefficients for that target radii should be a straight line. We checked this for all the stars in our sample and did not find such a constant, and so we conclude that the error introduced by a difference in the mean density is not the dominant source of difference in our inversion results. 

\subsection{Sun as a star} \label{appendix:mod_S}

\begin{deluxetable}{lcc}

\tablecaption{Reference Model using Degraded Solar Data} 
\label{tab:deg_sol} 
\tablehead{\colhead{Parameter} & \colhead{Unit} & \colhead{Value}} 

\startdata
$M$ &  $[\rm{M}_{\odot}]$ &  1.001 \\
$Y_{\rm{initial}}$ & \ldots &  0.282 \\
$Z_{\rm{initial}}$ & \ldots  &  0.021 \\
$L$ & $[L_{\odot}]$ & 1.057 \\
$T_{\rm{eff}}$ &  [K]  & 5849 \\
$R$ &  $[R_{\odot}]$ &  1.001 \\
$[$Fe/H$]$ & \ldots & 0.051
\enddata

\end{deluxetable}

\begin{figure} 
\centering
\plotone{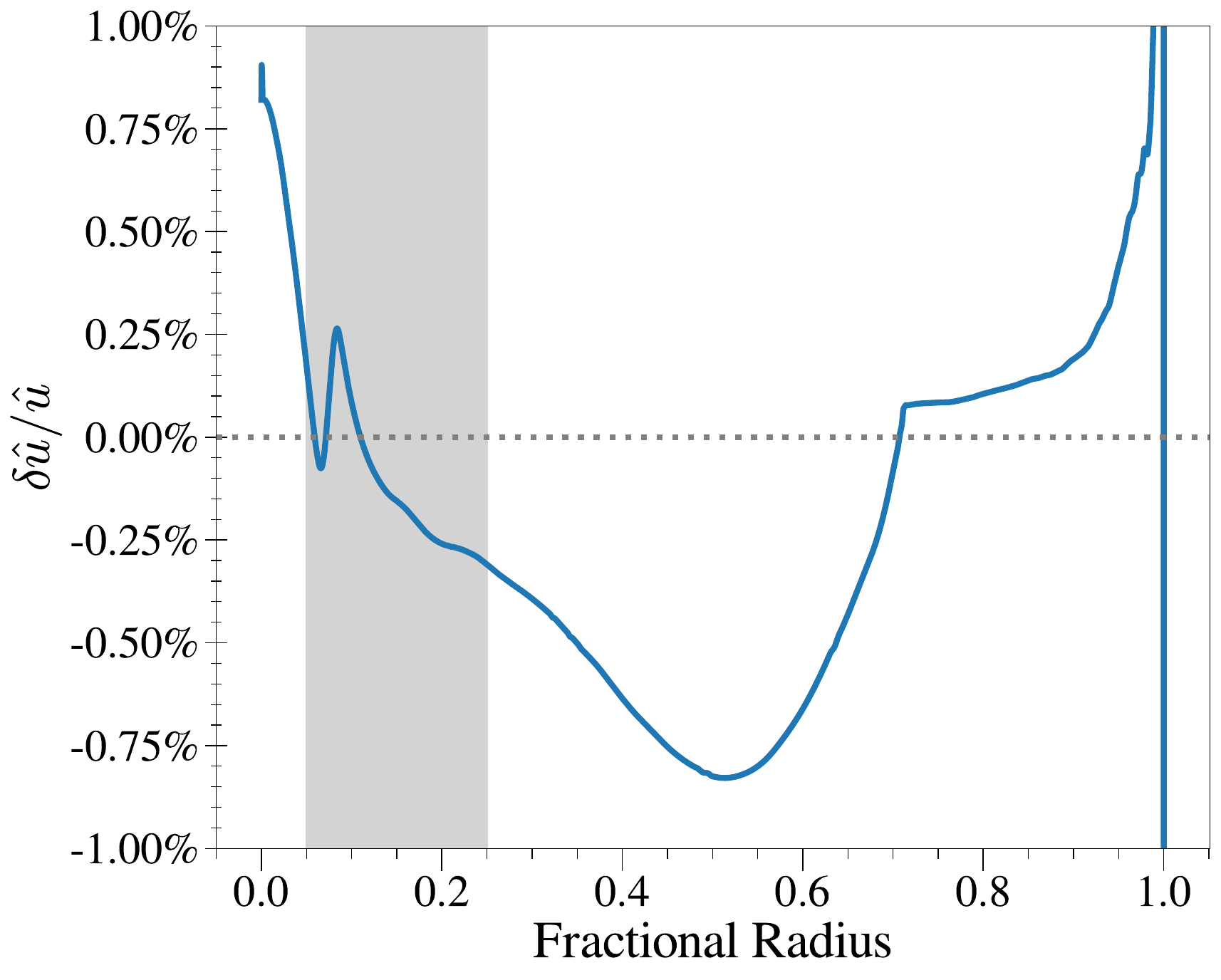} 
\caption{Relative difference in \(\hat{u}\) between our reference model, calculated from degraded solar data, and the calibrated solar model S \cite{1996Sci...272.1286C}. The shaded region shows the area that can be probed by structure inversions using only the reduced mode set of the degraded frequency data.} 
\label{fig:modS_comp}
\end{figure} 

\begin{figure} 
\centering

\plotone{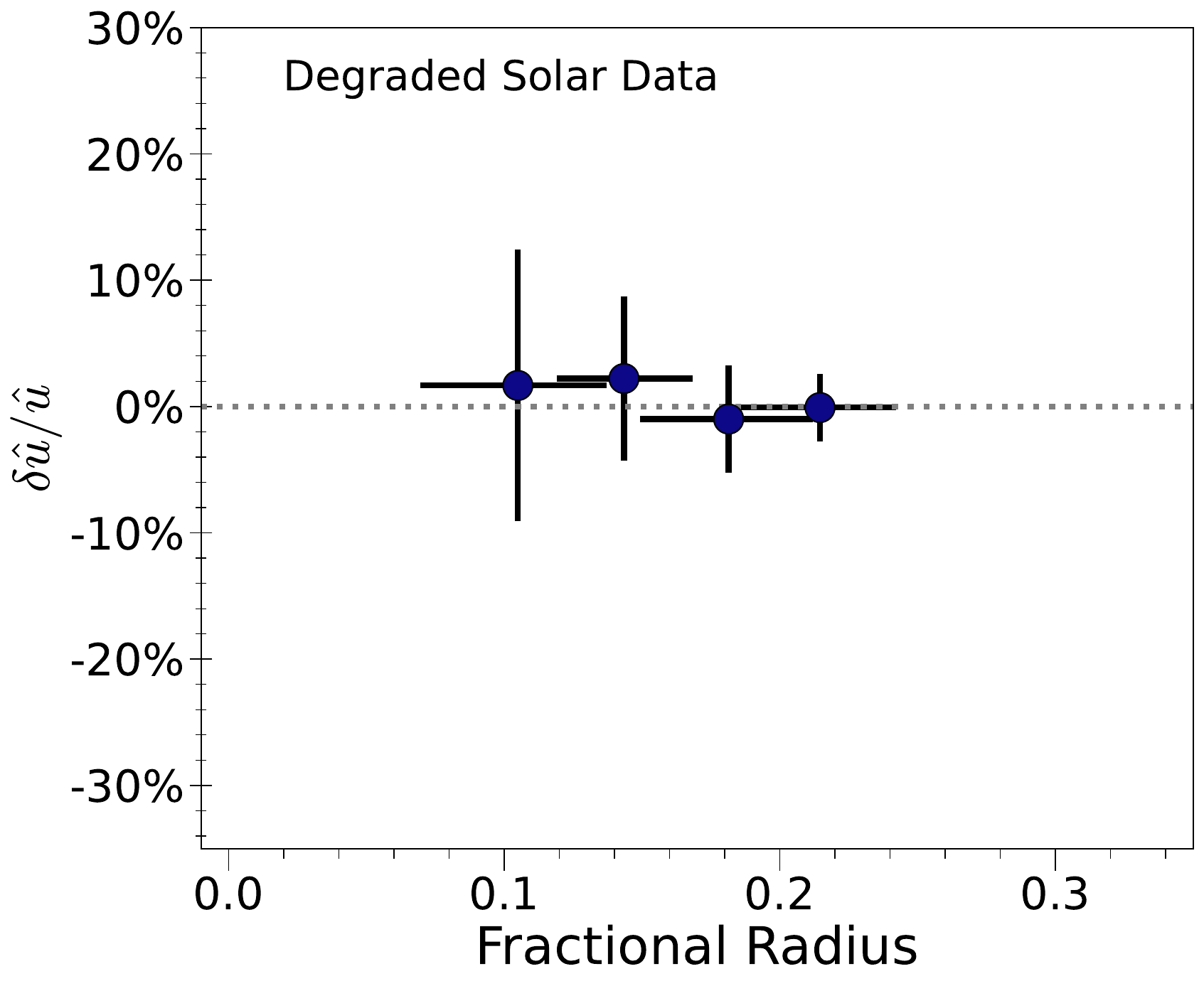} 
\caption{Inversion results of the degraded solar data. All symbols and colors have the same meaning as in Figure~\ref{fig:all_inv_res}.} 
\label{fig:inv_deg_solar}
\end{figure} 

In addition to the twelve target stars, we also obtain a reference model and structure inversion results using solar data that have been degraded to the level that was expected of results from \emph{Kepler} \citep[for details, see ][]{2017ApJ...835..172L}. Table~\ref{tab:deg_sol} lists the parameters of the reference model obtained with these data. The parameters of our model are comparable to those found across all the pipelines used in \citet{2017ApJ...835..172L}. To assess the quality of the fit in \(\hat{u}\), we compare the \(\hat{u}\) profile of our reference model to that of the calibrated standard solar model S of \cite{1996Sci...272.1286C}, shown in Figure~\rev{\ref{fig:modS_comp}}. Although we are not able to reproduce the full structure of a model calibrated with all the solar data, our reference model is a close match in the area probed by structure inversions using only low-degree modes. These differences are of the same order of magnitude as the differences inferred between model S and the Sun \citep{2009ApJ...699.1403B}. Thus, despite the limitations of the degraded data, we find a reference model sufficiently close for structure inversions. 

Using this reference model, we obtain suitable averaging kernels at four target radii and infer the difference in \(\hat{u}\) using the degraded solar frequencies as shown by Figure~\ref{fig:inv_deg_solar}. At all four target radii, our structure inversions show agreement within 1\(\sigma\) in \(\hat{u}\). Helioseismic inversions that use non-degraded solar data do show differences between the structure of the Sun and the structure of calibrated solar models \citep[e.g.,][]{2016LRSP...13....2B}; however, this results from using many more modes, with higher precision and at higher angular degrees, than are available for stars observed by \emph{Kepler}.

\section{Full Inversion Results} \label{appendix: Inversion details} 
\begin{figure*} [h]
\centering
\plotone{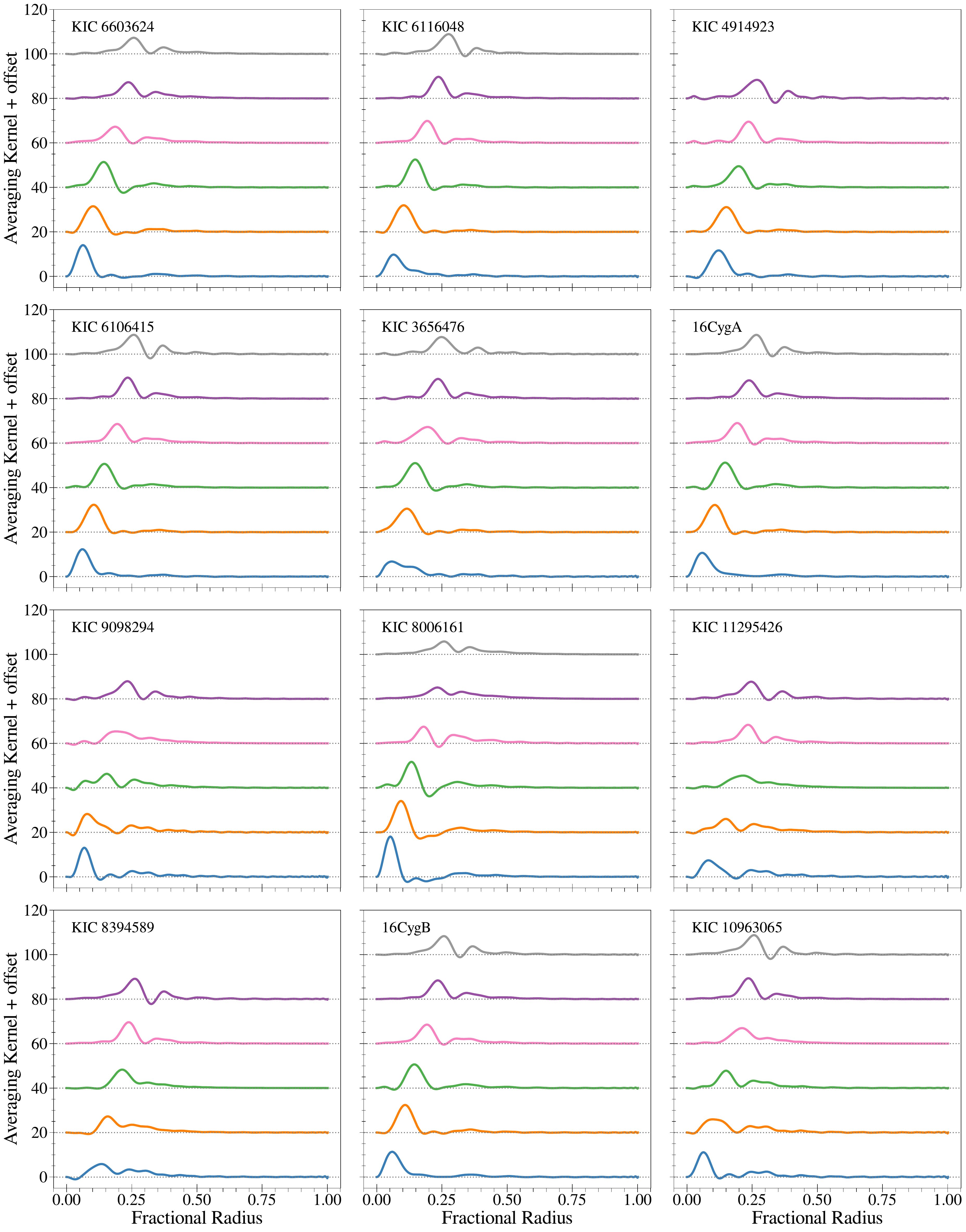} 
\caption{Averaging kernels \(\mathcal{K}\) for each of the target stars in our sample. For readability, the averaging kernels are offset, with a horizontal dotted line indicating the zero line for each kernel.}
\label{fig:avg_kern_all}
\end{figure*}

\begin{figure*} [h]
\centering
\plotone{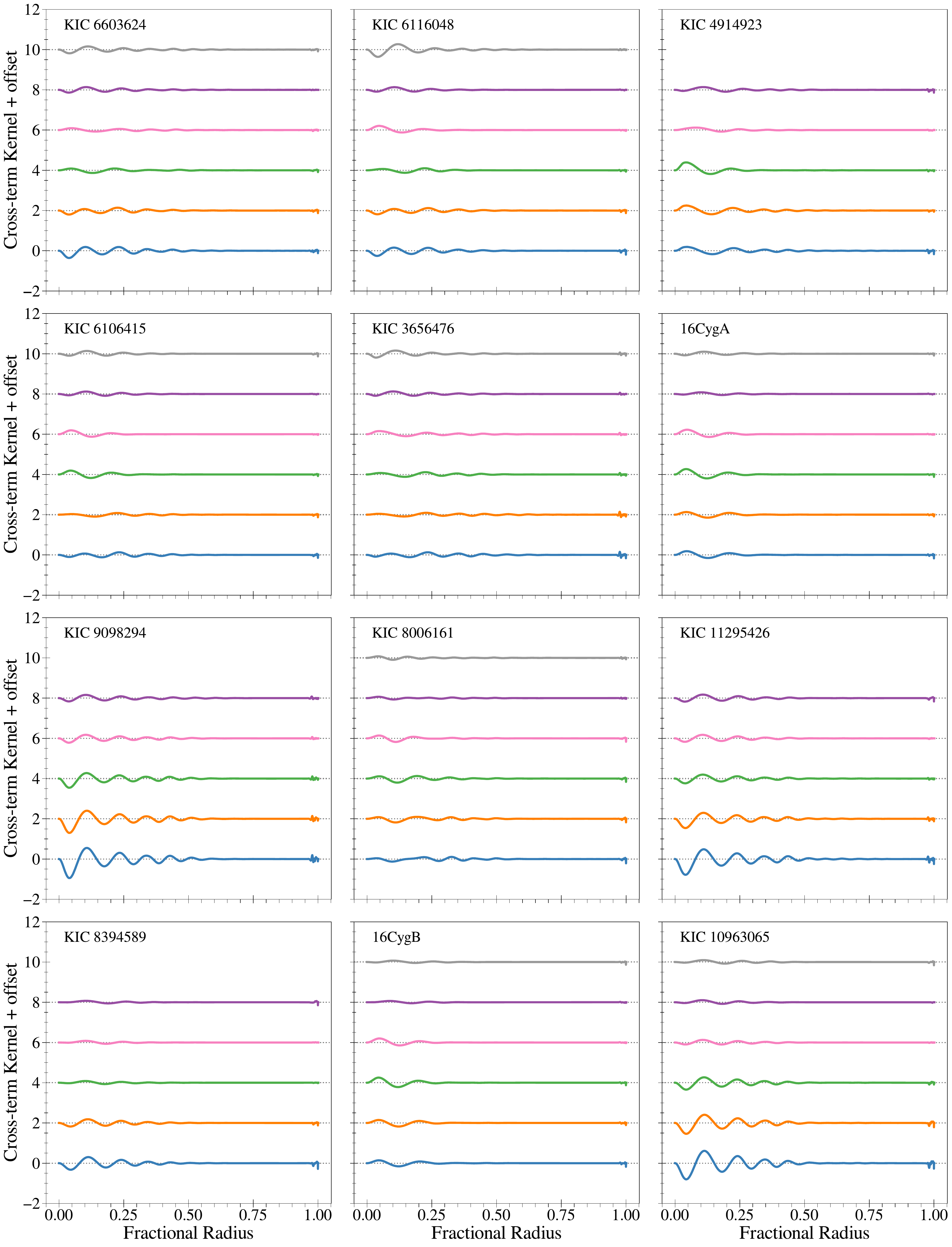} 
\caption{Cross-term kernels \(\mathcal{C}\) for each of the target stars in our sample. As with Figure~\ref{fig:avg_kern_all}, the kernels are offset for readability. Note that the y-axis scales differs from Figure~\ref{fig:avg_kern_all}.} 
\label{fig:cross_kern_all}
\end{figure*}

\begin{figure*} 
\centering
\plotone{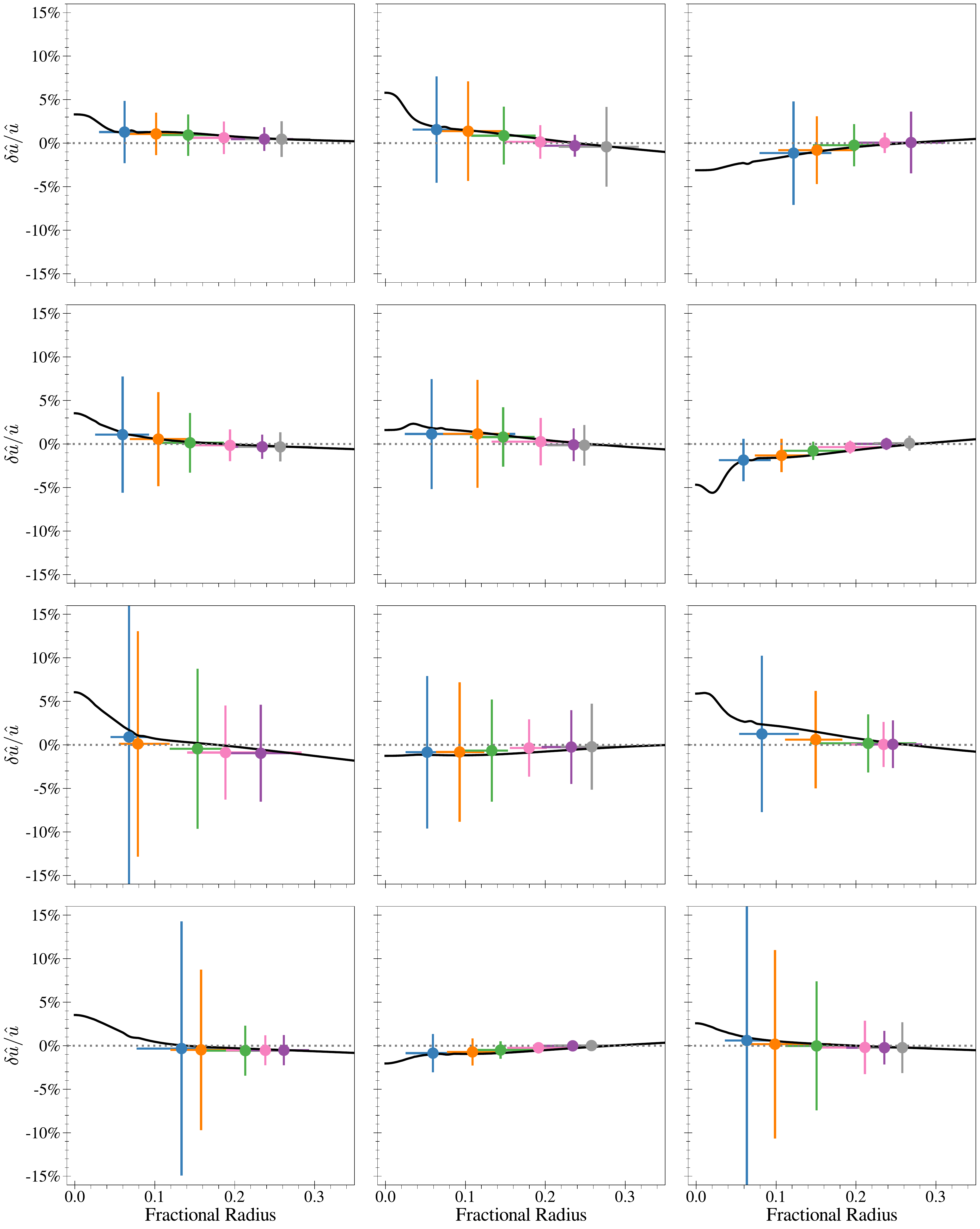} 
\caption{Results of using the averaging and cross term kernels shown in Figures \ref{fig:avg_kern_all} and \ref{fig:cross_kern_all} to recover the difference between the reference model and one of the additional models used to calibrate our inversion parameters. The black line indicates the true \(\delta \hat{u} / \hat{u}\) between the two models. The color of each point matches its corresponding averaging and cross term kernels in Figures \ref{fig:avg_kern_all} and \ref{fig:cross_kern_all}. }
\label{fig:mod-mod_inv} 
\end{figure*}

For each target star, we attempt a structure inversion at target radii of  \(r_{0}/R = 0.05, 0.10, 0.15, 0.20, 0.25, 0.30\). The target radius, however, is not necessarily the fractional radius where the averaging kernel is at its maximum value. We report the location of the maximum and FWHM of each averaging kernel and the \(\hat{u}\) values inferred at each target radius of each star in Table~\ref{tab:inv_res_full}.  Figures \ref{fig:avg_kern_all} and \ref{fig:cross_kern_all} show the averaging and cross-term kernels for these stars, respectively. \rev{We show in Figure~\ref{fig:mod-mod_inv} the results of model-model inversions between our reference model and one of the calibration models, as a test of the averaging kernel’s ability to recover a known difference. }

\begin{splitdeluxetable*}{cccccccBccccccc} 
\centering
\centerwidetable
\tablecaption{Location of the averaging kernel maximum (in fractional radius) and the corresponding FWHM and the infered dimensionless squared isothermal sound speed \(\hat{u}\), for each target radii $r_{0}/R = 0.05, 0.10, 0.15, 0.20, 0.25, 0.30$} 
\tablehead{\colhead{Star} & \colhead{$r_{\rm{max}}(0.05)$} & \colhead{$\hat{u}(0.05)$} & \colhead{$r_{\rm{max}}(0.10)$} & \colhead{$\hat{u}(0.10)$} & \colhead{$r_{\rm{max}}(0.15)$} & \colhead{$\hat{u}(0.15)$} & \colhead{Star} & \colhead{$r_{\rm{max}}(0.20)$} & \colhead{$\hat{u}(0.20)$}  & \colhead{$r_{\rm{max}}(0.25)$} & \colhead{$\hat{u}(0.25)$}   & \colhead{$r_{\rm{max}}(0.30)$}& \colhead{$\hat{u}(0.30)$}}
\label{tab:inv_res_full}

\startdata 
KIC 6603624 &  $0.062 ^{+0.029}_{-0.032}$ & $0.946 \pm 0.030$ &  $0.102 ^{+0.037}_{-0.035}$ & $0.924 \pm 0.021$ &  $0.142 ^{+0.034}_{-0.035}$ & $0.824 \pm 0.019$ & KIC 6603624 &  $0.187 ^{+0.036}_{-0.043}$ & $0.715 \pm 0.013$ &  $0.237 ^{+0.036}_{-0.042}$ & $0.618 \pm 0.008$ &  $0.259 ^{+0.036}_{-0.046}$ & $0.583 \pm 0.012$ \\
KIC 6116048 &  $0.064 ^{+0.039}_{-0.030}$ & $1.148 \pm 0.060$ &  $0.103 ^{+0.043}_{-0.039}$ & $1.071 \pm 0.055$ &  $0.148 ^{+0.040}_{-0.041}$ & $0.939 \pm 0.028$ & KIC 6116048 &  $0.194 ^{+0.041}_{-0.044}$ & $0.728 \pm 0.014$ &  $0.237 ^{+0.041}_{-0.043}$ & $0.647 \pm 0.008$ &  $0.276 ^{+0.040}_{-0.060}$ & $0.586 \pm 0.025$ \\
KIC 4914923 & \nodata & \nodata &  $0.122 ^{+0.048}_{-0.042}$ & $1.037 \pm 0.057$ &  $0.151 ^{+0.049}_{-0.048}$ & $0.905 \pm 0.034$ & KIC 4914923 &  $0.198 ^{+0.045}_{-0.052}$ & $0.739 \pm 0.018$ &  $0.236 ^{+0.043}_{-0.045}$ & $0.640 \pm 0.007$ &  $0.269 ^{+0.042}_{-0.069}$ & $0.580 \pm 0.019$ \\
KIC 6106415 &  $0.060 ^{+0.033}_{-0.035}$ & $1.020 \pm 0.063$ &  $0.104 ^{+0.039}_{-0.036}$ & $0.987 \pm 0.051$ &  $0.144 ^{+0.042}_{-0.039}$ & $0.841 \pm 0.029$ & KIC 6106415 &  $0.194 ^{+0.042}_{-0.044}$ & $0.713 \pm 0.013$ &  $0.235 ^{+0.041}_{-0.043}$ & $0.627 \pm 0.009$ &  $0.259 ^{+0.038}_{-0.055}$ & $0.583 \pm 0.009$ \\
KIC 3656476 &  $0.058 ^{+0.106}_{-0.034}$ & $0.862 \pm 0.061$ &  $0.116 ^{+0.040}_{-0.043}$ & $0.857 \pm 0.058$ &  $0.147 ^{+0.039}_{-0.042}$ & $0.808 \pm 0.029$ & KIC 3656476 &  $0.194 ^{+0.046}_{-0.062}$ & $0.704 \pm 0.020$ &  $0.235 ^{+0.036}_{-0.040}$ & $0.608 \pm 0.012$ &  $0.249 ^{+0.047}_{-0.047}$ & $0.576 \pm 0.014$ \\
16CygA &  $0.059 ^{+0.034}_{-0.031}$ & $0.930 \pm 0.022$ &  $0.107 ^{+0.035}_{-0.033}$ & $0.926 \pm 0.017$ &  $0.146 ^{+0.040}_{-0.037}$ & $0.837 \pm 0.009$ & 16CygA &  $0.193 ^{+0.035}_{-0.041}$ & $0.713 \pm 0.005$ &  $0.238 ^{+0.039}_{-0.040}$ & $0.612 \pm 0.004$ &  $0.267 ^{+0.033}_{-0.045}$ & $0.554 \pm 0.005$ \\
KIC 9098294 &  $0.068 ^{+0.023}_{-0.023}$ & $0.785 \pm 0.162$ &  $0.079 ^{+0.041}_{-0.024}$ & $0.826 \pm 0.121$ &  $0.154 ^{+0.032}_{-0.035}$ & $0.775 \pm 0.076$ & KIC 9098294 &  $0.189 ^{+0.094}_{-0.048}$ & $0.685 \pm 0.040$ &  $0.233 ^{+0.036}_{-0.046}$ & $0.548 \pm 0.035$ & \nodata & \nodata \\
KIC 8006161 &  $0.052 ^{+0.023}_{-0.027}$ & $0.802 \pm 0.065$ &  $0.093 ^{+0.031}_{-0.029}$ & $0.783 \pm 0.059$ &  $0.133 ^{+0.020}_{-0.037}$ & $0.753 \pm 0.041$ & KIC 8006161 &  $0.180 ^{+0.022}_{-0.024}$ & $0.659 \pm 0.021$ &  $0.232 ^{+0.035}_{-0.037}$ & $0.568 \pm 0.024$ &  $0.258 ^{+0.030}_{-0.039}$ & $0.524 \pm 0.026$ \\
KIC 11295426 & \nodata & \nodata &  $0.082 ^{+0.047}_{-0.028}$ & $0.912 \pm 0.084$ &  $0.150 ^{+0.034}_{-0.038}$ & $0.808 \pm 0.046$ & KIC 11295426 &  $0.215 ^{+0.060}_{-0.073}$ & $0.664 \pm 0.022$ &  $0.235 ^{+0.035}_{-0.041}$ & $0.632 \pm 0.016$ &  $0.246 ^{+0.037}_{-0.045}$ & $0.609 \pm 0.016$ \\
KIC 8394589 & \nodata & \nodata &  $0.134 ^{+0.051}_{-0.055}$ & $0.759 \pm 0.129$ &  $0.158 ^{+0.052}_{-0.038}$ & $0.747 \pm 0.075$ & KIC 8394589 &  $0.213 ^{+0.056}_{-0.054}$ & $0.645 \pm 0.019$ &  $0.238 ^{+0.037}_{-0.049}$ & $0.600 \pm 0.010$ &  $0.262 ^{+0.031}_{-0.054}$ & $0.550 \pm 0.010$ \\
16CygB &  $0.059 ^{+0.032}_{-0.034}$ & $0.848 \pm 0.019$ &  $0.109 ^{+0.034}_{-0.032}$ & $0.856 \pm 0.013$ &  $0.144 ^{+0.041}_{-0.035}$ & $0.801 \pm 0.008$ & 16CygB &  $0.193 ^{+0.035}_{-0.041}$ & $0.694 \pm 0.005$ &  $0.234 ^{+0.037}_{-0.041}$ & $0.603 \pm 0.003$ &  $0.258 ^{+0.035}_{-0.047}$ & $0.559 \pm 0.003$ \\
KIC 10963065 &  $0.063 ^{+0.027}_{-0.027}$ & $0.932 \pm 0.164$ &  $0.099 ^{+0.070}_{-0.041}$ & $0.927 \pm 0.105$ &  $0.151 ^{+0.038}_{-0.039}$ & $0.834 \pm 0.063$ & KIC 10963065 &  $0.211 ^{+0.057}_{-0.063}$ & $0.665 \pm 0.021$ &  $0.236 ^{+0.044}_{-0.048}$ & $0.616 \pm 0.012$ &  $0.258 ^{+0.036}_{-0.056}$ & $0.557 \pm 0.017$ \\
\enddata 

\end{splitdeluxetable*}

\clearpage
\bibliography{ms_inv_bib}{}

\begin{thebibliography}{}
\expandafter\ifx\csname natexlab\endcsname\relax\def\natexlab#1{#1}\fi
\providecommand{\url}[1]{\href{#1}{#1}}
\providecommand{\dodoi}[1]{doi:~\href{http://doi.org/#1}{\nolinkurl{#1}}}
\providecommand{\doeprint}[1]{\href{http://ascl.net/#1}{\nolinkurl{http://ascl.net/#1}}}
\providecommand{\doarXiv}[1]{\href{https://arxiv.org/abs/#1}{\nolinkurl{https://arxiv.org/abs/#1}}}

\bibitem[{{Aerts} {et~al.}(2010){Aerts}, {Christensen-Dalsgaard}, \&
  {Kurtz}}]{2010aste.book.....A}
{Aerts}, C., {Christensen-Dalsgaard}, J., \& {Kurtz}, D.~W. 2010,
  {Asteroseismology}, \dodoi{10.1007/978-1-4020-5803-5}

\bibitem[{{Aguirre B{\o}rsen-Koch} {et~al.}(2022){Aguirre B{\o}rsen-Koch},
  {R{\o}rsted}, {Justesen}, {Stokholm}, {Verma}, {Winther}, {Knudstrup},
  {Nielsen}, {Sahlholdt}, {Larsen}, {Cassisi}, {Serenelli}, {Casagrande},
  {Christensen-Dalsgaard}, {Davies}, {Ferguson}, {Lund}, {Weiss}, \&
  {White}}]{2022MNRAS.509.4344A}
{Aguirre B{\o}rsen-Koch}, V., {R{\o}rsted}, J.~L., {Justesen}, A.~B., {et~al.}
  2022, \mnras, 509, 4344, \dodoi{10.1093/mnras/stab2911}

\bibitem[{{Angelou} {et~al.}(2020){Angelou}, {Bellinger}, {Hekker}, {Mints},
  {Elsworth}, {Basu}, \& {Weiss}}]{2020MNRAS.493.4987A}
{Angelou}, G.~C., {Bellinger}, E.~P., {Hekker}, S., {et~al.} 2020, \mnras, 493,
  4987, \dodoi{10.1093/mnras/staa390}

\bibitem[{{Angulo} {et~al.}(1999){Angulo}, {Arnould}, {Rayet}, {Descouvemont},
  {Baye}, {Leclercq-Willain}, {Coc}, {Barhoumi}, {Aguer}, {Rolfs}, {Kunz},
  {Hammer}, {Mayer}, {Paradellis}, {Kossionides}, {Chronidou}, {Spyrou},
  {degl'Innocenti}, {Fiorentini}, {Ricci}, {Zavatarelli}, {Providencia},
  {Wolters}, {Soares}, {Grama}, {Rahighi}, {Shotter}, \& {Lamehi
  Rachti}}]{Angulo1999}
{Angulo}, C., {Arnould}, M., {Rayet}, M., {et~al.} 1999, \nphysa, 656, 3,
  \dodoi{10.1016/S0375-9474(99)00030-5}

\bibitem[{{Backus} \& {Gilbert}(1968)}]{1968GeoJ...16..169B}
{Backus}, G., \& {Gilbert}, F. 1968, Geophysical Journal, 16, 169,
  \dodoi{10.1111/j.1365-246X.1968.tb00216.x}

\bibitem[{{Backus} \& {Gilbert}(1970)}]{1970RSPTA.266..123B}
---. 1970, Philosophical Transactions of the Royal Society of London Series A,
  266, 123, \dodoi{10.1098/rsta.1970.0005}

\bibitem[{{Ball} \& {Gizon}(2014)}]{2014A&A...568A.123B}
{Ball}, W.~H., \& {Gizon}, L. 2014, \aap, 568, A123,
  \dodoi{10.1051/0004-6361/201424325}

\bibitem[{{Basu}(2003)}]{2003Ap&SS.284..153B}
{Basu}, S. 2003, \apss, 284, 153

\bibitem[{{Basu}(2016)}]{2016LRSP...13....2B}
---. 2016, Living Reviews in Solar Physics, 13, 2,
  \dodoi{10.1007/s41116-016-0003-4}

\bibitem[{{Basu} \& {Chaplin}(2017)}]{2017asda.book.....B}
{Basu}, S., \& {Chaplin}, W.~J. 2017, {Asteroseismic Data Analysis: Foundations
  and Techniques}

\bibitem[{{Basu} {et~al.}(2009){Basu}, {Chaplin}, {Elsworth}, {New}, \&
  {Serenelli}}]{2009ApJ...699.1403B}
{Basu}, S., {Chaplin}, W.~J., {Elsworth}, Y., {New}, R., \& {Serenelli}, A.~M.
  2009, \apj, 699, 1403, \dodoi{10.1088/0004-637X/699/2/1403}

\bibitem[{{Basu} \& {Christensen-Dalsgaard}(1997)}]{1997A&A...322L...5B}
{Basu}, S., \& {Christensen-Dalsgaard}, J. 1997, \aap, 322, L5.
\newblock \doarXiv{astro-ph/9702162}

\bibitem[{{Bazot} {et~al.}(2008){Bazot}, {Bourguignon}, \&
  {Christensen-Dalsgaard}}]{2008MmSAI..79..660B}
{Bazot}, M., {Bourguignon}, S., \& {Christensen-Dalsgaard}, J. 2008, \memsai,
  79, 660, \dodoi{10.48550/arXiv.0803.2529}

\bibitem[{{Bellinger} {et~al.}(2016){Bellinger}, {Angelou}, {Hekker}, {Basu},
  {Ball}, \& {Guggenberger}}]{2016ApJ...830...31B}
{Bellinger}, E.~P., {Angelou}, G.~C., {Hekker}, S., {et~al.} 2016, \apj, 830,
  31, \dodoi{10.3847/0004-637X/830/1/31}

\bibitem[{{Bellinger} {et~al.}(2020{\natexlab{a}}){Bellinger}, {Basu}, \&
  {Hekker}}]{2020ASSP...57..171B}
{Bellinger}, E.~P., {Basu}, S., \& {Hekker}, S. 2020{\natexlab{a}}, in
  Astrophysics and Space Science Proceedings, Vol.~57, Dynamics of the Sun and
  Stars; Honoring the Life and Work of Michael J. Thompson, ed. M.~J.~P.~F.~G.
  {Monteiro}, R.~A. {Garc{\'\i}a}, J.~{Christensen-Dalsgaard}, \& S.~W.
  {McIntosh}, 171--183, \dodoi{10.1007/978-3-030-55336-4_25}

\bibitem[{{Bellinger} {et~al.}(2017){Bellinger}, {Basu}, {Hekker}, \&
  {Ball}}]{2017ApJ...851...80B}
{Bellinger}, E.~P., {Basu}, S., {Hekker}, S., \& {Ball}, W.~H. 2017, \apj, 851,
  80, \dodoi{10.3847/1538-4357/aa9848}

\bibitem[{{Bellinger} {et~al.}(2019{\natexlab{a}}){Bellinger}, {Basu},
  {Hekker}, \& {Christensen-Dalsgaard}}]{2019ApJ...885..143B}
{Bellinger}, E.~P., {Basu}, S., {Hekker}, S., \& {Christensen-Dalsgaard}, J.
  2019{\natexlab{a}}, \apj, 885, 143, \dodoi{10.3847/1538-4357/ab4a0d}

\bibitem[{{Bellinger} {et~al.}(2021){Bellinger}, {Basu}, {Hekker},
  {Christensen-Dalsgaard}, \& {Ball}}]{2021ApJ...915..100B}
{Bellinger}, E.~P., {Basu}, S., {Hekker}, S., {Christensen-Dalsgaard}, J., \&
  {Ball}, W.~H. 2021, \apj, 915, 100, \dodoi{10.3847/1538-4357/ac0051}

\bibitem[{{Bellinger} \& {Christensen-Dalsgaard}(2019)}]{2019ApJ...887L...1B}
{Bellinger}, E.~P., \& {Christensen-Dalsgaard}, J. 2019, \apjl, 887, L1,
  \dodoi{10.3847/2041-8213/ab43e7}

\bibitem[{{Bellinger} {et~al.}(2019{\natexlab{b}}){Bellinger}, {Hekker},
  {Angelou}, {Stokholm}, \& {Basu}}]{2019A&A...622A.130B}
{Bellinger}, E.~P., {Hekker}, S., {Angelou}, G.~C., {Stokholm}, A., \& {Basu},
  S. 2019{\natexlab{b}}, \aap, 622, A130, \dodoi{10.1051/0004-6361/201834461}

\bibitem[{{Bellinger} {et~al.}(2020{\natexlab{b}}){Bellinger}, {Kanbur},
  {Bhardwaj}, \& {Marconi}}]{2020MNRAS.491.4752B}
{Bellinger}, E.~P., {Kanbur}, S.~M., {Bhardwaj}, A., \& {Marconi}, M.
  2020{\natexlab{b}}, \mnras, 491, 4752, \dodoi{10.1093/mnras/stz3292}

\bibitem[{{Borucki} {et~al.}(2010){Borucki}, {Koch}, {Basri}, {Batalha},
  {Brown}, {Caldwell}, {Caldwell}, {Christensen-Dalsgaard}, {Cochran},
  {DeVore}, {Dunham}, {Dupree}, {Gautier}, {Geary}, {Gilliland}, {Gould},
  {Howell}, {Jenkins}, {Kondo}, {Latham}, {Marcy}, {Meibom}, {Kjeldsen},
  {Lissauer}, {Monet}, {Morrison}, {Sasselov}, {Tarter}, {Boss}, {Brownlee},
  {Owen}, {Buzasi}, {Charbonneau}, {Doyle}, {Fortney}, {Ford}, {Holman},
  {Seager}, {Steffen}, {Welsh}, {Rowe}, {Anderson}, {Buchhave}, {Ciardi},
  {Walkowicz}, {Sherry}, {Horch}, {Isaacson}, {Everett}, {Fischer}, {Torres},
  {Johnson}, {Endl}, {MacQueen}, {Bryson}, {Dotson}, {Haas}, {Kolodziejczak},
  {Van Cleve}, {Chandrasekaran}, {Twicken}, {Quintana}, {Clarke}, {Allen},
  {Li}, {Wu}, {Tenenbaum}, {Verner}, {Bruhweiler}, {Barnes}, \&
  {Prsa}}]{2010Sci...327..977B}
{Borucki}, W.~J., {Koch}, D., {Basri}, G., {et~al.} 2010, Science, 327, 977,
  \dodoi{10.1126/science.1185402}

\bibitem[{{Buldgen} {et~al.}(2022{\natexlab{a}}){Buldgen}, {B{\'e}trisey},
  {Roxburgh}, {Vorontsov}, \& {Reese}}]{2022FrASS...9.2373B}
{Buldgen}, G., {B{\'e}trisey}, J., {Roxburgh}, I.~W., {Vorontsov}, S.~V., \&
  {Reese}, D.~R. 2022{\natexlab{a}}, Frontiers in Astronomy and Space Sciences,
  9, 942373, \dodoi{10.3389/fspas.2022.942373}

\bibitem[{{Buldgen} {et~al.}(2022{\natexlab{b}}){Buldgen}, {Farnir},
  {Eggenberger}, {B{\'e}trisey}, {Pezzotti}, {Pin{\c{c}}on}, {Deal}, \&
  {Salmon}}]{2022A&A...661A.143B}
{Buldgen}, G., {Farnir}, M., {Eggenberger}, P., {et~al.} 2022{\natexlab{b}},
  \aap, 661, A143, \dodoi{10.1051/0004-6361/202142001}

\bibitem[{{Buldgen} {et~al.}(2015){Buldgen}, {Reese}, \&
  {Dupret}}]{2015A&A...583A..62B}
{Buldgen}, G., {Reese}, D.~R., \& {Dupret}, M.~A. 2015, \aap, 583, A62,
  \dodoi{10.1051/0004-6361/201526390}

\bibitem[{{Buldgen} {et~al.}(2017){Buldgen}, {Reese}, \&
  {Dupret}}]{2017A&A...598A..21B}
---. 2017, \aap, 598, A21, \dodoi{10.1051/0004-6361/201629485}

\bibitem[{{Chaplin} \& {Miglio}(2013)}]{2013ARA&A..51..353C}
{Chaplin}, W.~J., \& {Miglio}, A. 2013, \araa, 51, 353,
  \dodoi{10.1146/annurev-astro-082812-140938}

\bibitem[{{Charpinet} {et~al.}(2005){Charpinet}, {Fontaine}, {Brassard},
  {Green}, \& {Chayer}}]{2005A&A...437..575C}
{Charpinet}, S., {Fontaine}, G., {Brassard}, P., {Green}, E.~M., \& {Chayer},
  P. 2005, \aap, 437, 575, \dodoi{10.1051/0004-6361:20052709}

\bibitem[{{Christensen-Dalsgaard}(2021)}]{2021LRSP...18....2C}
{Christensen-Dalsgaard}, J. 2021, Living Reviews in Solar Physics, 18, 2,
  \dodoi{10.1007/s41116-020-00028-3}

\bibitem[{{Christensen-Dalsgaard} {et~al.}(1996){Christensen-Dalsgaard},
  {Dappen}, {Ajukov}, {Anderson}, {Antia}, {Basu}, {Baturin}, {Berthomieu},
  {Chaboyer}, {Chitre}, {Cox}, {Demarque}, {Donatowicz}, {Dziembowski},
  {Gabriel}, {Gough}, {Guenther}, {Guzik}, {Harvey}, {Hill}, {Houdek},
  {Iglesias}, {Kosovichev}, {Leibacher}, {Morel}, {Proffitt}, {Provost},
  {Reiter}, {Rhodes}, {Rogers}, {Roxburgh}, {Thompson}, \&
  {Ulrich}}]{1996Sci...272.1286C}
{Christensen-Dalsgaard}, J., {Dappen}, W., {Ajukov}, S.~V., {et~al.} 1996,
  Science, 272, 1286, \dodoi{10.1126/science.272.5266.1286}

\bibitem[{{Chugunov} {et~al.}(2007){Chugunov}, {Dewitt}, \&
  {Yakovlev}}]{Chugunov2007}
{Chugunov}, A.~I., {Dewitt}, H.~E., \& {Yakovlev}, D.~G. 2007, \prd, 76,
  025028, \dodoi{10.1103/PhysRevD.76.025028}

\bibitem[{{Cox} \& {Giuli}(1968)}]{1968pss..book.....C}
{Cox}, J.~P., \& {Giuli}, R.~T. 1968, {Principles of stellar structure}

\bibitem[{{Creevey} {et~al.}(2023){Creevey}, {Sordo}, {Pailler}, {Fr{\'e}mat},
  {Heiter}, {Th{\'e}venin}, {Andrae}, {Fouesneau}, {Lobel}, {Bailer-Jones},
  {Garabato}, {Bellas-Velidis}, {Brugaletta}, {Lorca}, {Ordenovic}, {Palicio},
  {Sarro}, {Delchambre}, {Drimmel}, {Rybizki}, {Torralba Elipe}, {Korn},
  {Recio-Blanco}, {Schultheis}, {De Angeli}, {Montegriffo}, {Abreu Aramburu},
  {Accart}, {{\'A}lvarez}, {Bakker}, {Brouillet}, {Burlacu}, {Carballo},
  {Casamiquela}, {Chiavassa}, {Contursi}, {Cooper}, {Dafonte}, {Dapergolas},
  {de Laverny}, {Dharmawardena}, {Edvardsson}, {Le Fustec},
  {Garc{\'\i}a-Lario}, {Garc{\'\i}a-Torres}, {Gomez},
  {Gonz{\'a}lez-Santamar{\'\i}a}, {Hatzidimitriou}, {Jean-Antoine Piccolo},
  {Kontiza}, {Kordopatis}, {Lanzafame}, {Lebreton}, {Licata}, {Lindstr{\o}m},
  {Livanou}, {Magdaleno Romeo}, {Manteiga}, {Marocco}, {Marshall}, {Mary},
  {Nicolas}, {Pallas-Quintela}, {Panem}, {Pichon}, {Poggio}, {Riclet}, {Robin},
  {Santove{\~n}a}, {Silvelo}, {Slezak}, {Smart}, {Soubiran}, {S{\"u}veges},
  {Ulla}, {Utrilla}, {Vallenari}, {Zhao}, {Zorec}, {Barrado}, {Bijaoui},
  {Bouret}, {Blomme}, {Brott}, {Cassisi}, {Kochukhov}, {Martayan}, {Shulyak},
  \& {Silvester}}]{2023A&A...674A..26C}
{Creevey}, O.~L., {Sordo}, R., {Pailler}, F., {et~al.} 2023, \aap, 674, A26,
  \dodoi{10.1051/0004-6361/202243688}

\bibitem[{{Cyburt} {et~al.}(2010){Cyburt}, {Amthor}, {Ferguson}, {Meisel},
  {Smith}, {Warren}, {Heger}, {Hoffman}, {Rauscher}, {Sakharuk}, {Schatz},
  {Thielemann}, \& {Wiescher}}]{Cyburt2010}
{Cyburt}, R.~H., {Amthor}, A.~M., {Ferguson}, R., {et~al.} 2010, \apjs, 189,
  240, \dodoi{10.1088/0067-0049/189/1/240}

\bibitem[{{Davies} {et~al.}(2016){Davies}, {Silva Aguirre}, {Bedding},
  {Handberg}, {Lund}, {Chaplin}, {Huber}, {White}, {Benomar}, {Hekker}, {Basu},
  {Campante}, {Christensen-Dalsgaard}, {Elsworth}, {Karoff}, {Kjeldsen},
  {Lundkvist}, {Metcalfe}, \& {Stello}}]{2016MNRAS.456.2183D}
{Davies}, G.~R., {Silva Aguirre}, V., {Bedding}, T.~R., {et~al.} 2016, \mnras,
  456, 2183, \dodoi{10.1093/mnras/stv2593}

\bibitem[{{Dziembowski} {et~al.}(1990){Dziembowski}, {Pamyatnykh}, \&
  {Sienkiewicz}}]{1990MNRAS.244..542D}
{Dziembowski}, W.~A., {Pamyatnykh}, A.~A., \& {Sienkiewicz}, R. 1990, \mnras,
  244, 542

\bibitem[{{Elliott}(1996)}]{1996MNRAS.280.1244E}
{Elliott}, J.~R. 1996, \mnras, 280, 1244, \dodoi{10.1093/mnras/280.4.1244}

\bibitem[{{Ferguson} {et~al.}(2005){Ferguson}, {Alexander}, {Allard}, {Barman},
  {Bodnarik}, {Hauschildt}, {Heffner-Wong}, \& {Tamanai}}]{Ferguson2005}
{Ferguson}, J.~W., {Alexander}, D.~R., {Allard}, F., {et~al.} 2005, \apj, 623,
  585, \dodoi{10.1086/428642}

\bibitem[{{Frandsen} {et~al.}(2002){Frandsen}, {Carrier}, {Aerts}, {Stello},
  {Maas}, {Burnet}, {Bruntt}, {Teixeira}, {de Medeiros}, {Bouchy}, {Kjeldsen},
  {Pijpers}, \& {Christensen-Dalsgaard}}]{2002A&A...394L...5F}
{Frandsen}, S., {Carrier}, F., {Aerts}, C., {et~al.} 2002, \aap, 394, L5,
  \dodoi{10.1051/0004-6361:20021281}

\bibitem[{{Fuller} {et~al.}(1985){Fuller}, {Fowler}, \& {Newman}}]{Fuller1985}
{Fuller}, G.~M., {Fowler}, W.~A., \& {Newman}, M.~J. 1985, \apj, 293, 1,
  \dodoi{10.1086/163208}

\bibitem[{{Furlan} {et~al.}(2018){Furlan}, {Ciardi}, {Cochran}, {Everett},
  {Latham}, {Marcy}, {Buchhave}, {Endl}, {Isaacson}, {Petigura}, {Gautier},
  {Huber}, {Bieryla}, {Borucki}, {Brugamyer}, {Caldwell}, {Cochran}, {Howard},
  {Howell}, {Johnson}, {MacQueen}, {Quinn}, {Robertson}, {Mathur}, \&
  {Batalha}}]{2018ApJ...861..149F}
{Furlan}, E., {Ciardi}, D.~R., {Cochran}, W.~D., {et~al.} 2018, \apj, 861, 149,
  \dodoi{10.3847/1538-4357/aaca34}

\bibitem[{{Gaia Collaboration} {et~al.}(2016){Gaia Collaboration}, {Prusti},
  {de Bruijne}, {Brown}, {Vallenari}, {Babusiaux}, {Bailer-Jones}, {Bastian},
  {Biermann}, {Evans}, {Eyer}, {Jansen}, {Jordi}, {Klioner}, {Lammers},
  {Lindegren}, {Luri}, {Mignard}, {Milligan}, {Panem}, {Poinsignon},
  {Pourbaix}, {Randich}, {Sarri}, {Sartoretti}, {Siddiqui}, {Soubiran},
  {Valette}, {van Leeuwen}, {Walton}, {Aerts}, {Arenou}, {Cropper}, {Drimmel},
  {H{\o}g}, {Katz}, {Lattanzi}, {O'Mullane}, {Grebel}, {Holland}, {Huc},
  {Passot}, {Bramante}, {Cacciari}, {Casta{\~n}eda}, {Chaoul}, {Cheek}, {De
  Angeli}, {Fabricius}, {Guerra}, {Hern{\'a}ndez}, {Jean-Antoine-Piccolo},
  {Masana}, {Messineo}, {Mowlavi}, {Nienartowicz}, {Ord{\'o}{\~n}ez-Blanco},
  {Panuzzo}, {Portell}, {Richards}, {Riello}, {Seabroke}, {Tanga},
  {Th{\'e}venin}, {Torra}, {Els}, {Gracia-Abril}, {Comoretto},
  {Garcia-Reinaldos}, {Lock}, {Mercier}, {Altmann}, {Andrae}, {Astraatmadja},
  {Bellas-Velidis}, {Benson}, {Berthier}, {Blomme}, {Busso}, {Carry},
  {Cellino}, {Clementini}, {Cowell}, {Creevey}, {Cuypers}, {Davidson}, {De
  Ridder}, {de Torres}, {Delchambre}, {Dell'Oro}, {Ducourant}, {Fr{\'e}mat},
  {Garc{\'\i}a-Torres}, {Gosset}, {Halbwachs}, {Hambly}, {Harrison}, {Hauser},
  {Hestroffer}, {Hodgkin}, {Huckle}, {Hutton}, {Jasniewicz}, {Jordan},
  {Kontizas}, {Korn}, {Lanzafame}, {Manteiga}, {Moitinho}, {Muinonen},
  {Osinde}, {Pancino}, {Pauwels}, {Petit}, {Recio-Blanco}, {Robin}, {Sarro},
  {Siopis}, {Smith}, {Smith}, {Sozzetti}, {Thuillot}, {van Reeven}, {Viala},
  {Abbas}, {Abreu Aramburu}, {Accart}, {Aguado}, {Allan}, {Allasia},
  {Altavilla}, {{\'A}lvarez}, {Alves}, {Anderson}, {Andrei}, {Anglada Varela},
  {Antiche}, {Antoja}, {Ant{\'o}n}, {Arcay}, {Atzei}, {Ayache}, {Bach},
  {Baker}, {Balaguer-N{\'u}{\~n}ez}, {Barache}, {Barata}, {Barbier}, {Barblan},
  {Baroni}, {Barrado y Navascu{\'e}s}, {Barros}, {Barstow}, {Becciani},
  {Bellazzini}, {Bellei}, {Bello Garc{\'\i}a}, {Belokurov}, {Bendjoya},
  {Berihuete}, {Bianchi}, {Bienaym{\'e}}, {Billebaud}, {Blagorodnova},
  {Blanco-Cuaresma}, {Boch}, {Bombrun}, {Borrachero}, {Bouquillon}, {Bourda},
  {Bouy}, {Bragaglia}, {Breddels}, {Brouillet}, {Br{\"u}semeister},
  {Bucciarelli}, {Budnik}, {Burgess}, {Burgon}, {Burlacu}, {Busonero}, {Buzzi},
  {Caffau}, {Cambras}, {Campbell}, {Cancelliere}, {Cantat-Gaudin}, {Carlucci},
  {Carrasco}, {Castellani}, {Charlot}, {Charnas}, {Charvet}, {Chassat},
  {Chiavassa}, {Clotet}, {Cocozza}, {Collins}, {Collins}, {Costigan}, {Crifo},
  {Cross}, {Crosta}, {Crowley}, {Dafonte}, {Damerdji}, {Dapergolas}, {David},
  {David}, {De Cat}, {de Felice}, {de Laverny}, {De Luise}, {De March}, {de
  Martino}, {de Souza}, {Debosscher}, {del Pozo}, {Delbo}, {Delgado},
  {Delgado}, {di Marco}, {Di Matteo}, {Diakite}, {Distefano}, {Dolding}, {Dos
  Anjos}, {Drazinos}, {Dur{\'a}n}, {Dzigan}, {Ecale}, {Edvardsson}, {Enke},
  {Erdmann}, {Escolar}, {Espina}, {Evans}, {Eynard Bontemps}, {Fabre},
  {Fabrizio}, {Faigler}, {Falc{\~a}o}, {Farr{\`a}s Casas}, {Faye}, {Federici},
  {Fedorets}, {Fern{\'a}ndez-Hern{\'a}ndez}, {Fernique}, {Fienga}, {Figueras},
  {Filippi}, {Findeisen}, {Fonti}, {Fouesneau}, {Fraile}, {Fraser}, {Fuchs},
  {Furnell}, {Gai}, {Galleti}, {Galluccio}, {Garabato}, {Garc{\'\i}a-Sedano},
  {Gar{\'e}}, {Garofalo}, {Garralda}, {Gavras}, {Gerssen}, {Geyer}, {Gilmore},
  {Girona}, {Giuffrida}, {Gomes}, {Gonz{\'a}lez-Marcos},
  {Gonz{\'a}lez-N{\'u}{\~n}ez}, {Gonz{\'a}lez-Vidal}, {Granvik}, {Guerrier},
  {Guillout}, {Guiraud}, {G{\'u}rpide}, {Guti{\'e}rrez-S{\'a}nchez}, {Guy},
  {Haigron}, {Hatzidimitriou}, {Haywood}, {Heiter}, {Helmi}, {Hobbs},
  {Hofmann}, {Holl}, {Holland}, {Hunt}, {Hypki}, {Icardi}, {Irwin}, {Jevardat
  de Fombelle}, {Jofr{\'e}}, {Jonker}, {Jorissen}, {Julbe}, {Karampelas},
  {Kochoska}, {Kohley}, {Kolenberg}, {Kontizas}, {Koposov}, {Kordopatis},
  {Koubsky}, {Kowalczyk}, {Krone-Martins}, {Kudryashova}, {Kull}, {Bachchan},
  {Lacoste-Seris}, {Lanza}, {Lavigne}, {Le Poncin-Lafitte}, {Lebreton},
  {Lebzelter}, {Leccia}, {Leclerc}, {Lecoeur-Taibi}, {Lemaitre}, {Lenhardt},
  {Leroux}, {Liao}, {Licata}, {Lindstr{\o}m}, {Lister}, {Livanou}, {Lobel},
  {L{\"o}ffler}, {L{\'o}pez}, {Lopez-Lozano}, {Lorenz}, {Loureiro},
  {MacDonald}, {Magalh{\~a}es Fernandes}, {Managau}, {Mann}, {Mantelet},
  {Marchal}, {Marchant}, {Marconi}, {Marie}, {Marinoni}, {Marrese},
  {Marschalk{\'o}}, {Marshall}, {Mart{\'\i}n-Fleitas}, {Martino}, {Mary},
  {Matijevi{\v{c}}}, {Mazeh}, {McMillan}, {Messina}, {Mestre}, {Michalik},
  {Millar}, {Miranda}, {Molina}, {Molinaro}, {Molinaro}, {Moln{\'a}r},
  {Moniez}, {Montegriffo}, {Monteiro}, {Mor}, {Mora}, {Morbidelli}, {Morel},
  {Morgenthaler}, {Morley}, {Morris}, {Mulone}, {Muraveva}, {Musella},
  {Narbonne}, {Nelemans}, {Nicastro}, {Noval}, {Ord{\'e}novic},
  {Ordieres-Mer{\'e}}, {Osborne}, {Pagani}, {Pagano}, {Pailler}, {Palacin},
  {Palaversa}, {Parsons}, {Paulsen}, {Pecoraro}, {Pedrosa}, {Pentik{\"a}inen},
  {Pereira}, {Pichon}, {Piersimoni}, {Pineau}, {Plachy}, {Plum}, {Poujoulet},
  {Pr{\v{s}}a}, {Pulone}, {Ragaini}, {Rago}, {Rambaux}, {Ramos-Lerate},
  {Ranalli}, {Rauw}, {Read}, {Regibo}, {Renk}, {Reyl{\'e}}, {Ribeiro},
  {Rimoldini}, {Ripepi}, {Riva}, {Rixon}, {Roelens}, {Romero-G{\'o}mez},
  {Rowell}, {Royer}, {Rudolph}, {Ruiz-Dern}, {Sadowski}, {Sagrist{\`a}
  Sell{\'e}s}, {Sahlmann}, {Salgado}, {Salguero}, {Sarasso}, {Savietto},
  {Schnorhk}, {Schultheis}, {Sciacca}, {Segol}, {Segovia}, {Segransan},
  {Serpell}, {Shih}, {Smareglia}, {Smart}, {Smith}, {Solano}, {Solitro},
  {Sordo}, {Soria Nieto}, {Souchay}, {Spagna}, {Spoto}, {Stampa}, {Steele},
  {Steidelm{\"u}ller}, {Stephenson}, {Stoev}, {Suess}, {S{\"u}veges}, {Surdej},
  {Szabados}, {Szegedi-Elek}, {Tapiador}, {Taris}, {Tauran}, {Taylor},
  {Teixeira}, {Terrett}, {Tingley}, {Trager}, {Turon}, {Ulla}, {Utrilla},
  {Valentini}, {van Elteren}, {Van Hemelryck}, {van Leeuwen}, {Varadi},
  {Vecchiato}, {Veljanoski}, {Via}, {Vicente}, {Vogt}, {Voss}, {Votruba},
  {Voutsinas}, {Walmsley}, {Weiler}, {Weingrill}, {Werner}, {Wevers},
  {Whitehead}, {Wyrzykowski}, {Yoldas}, {{\v{Z}}erjal}, {Zucker}, {Zurbach},
  {Zwitter}, {Alecu}, {Allen}, {Allende Prieto}, {Amorim},
  {Anglada-Escud{\'e}}, {Arsenijevic}, {Azaz}, {Balm}, {Beck}, {Bernstein},
  {Bigot}, {Bijaoui}, {Blasco}, {Bonfigli}, {Bono}, {Boudreault}, {Bressan},
  {Brown}, {Brunet}, {Bunclark}, {Buonanno}, {Butkevich}, {Carret}, {Carrion},
  {Chemin}, {Ch{\'e}reau}, {Corcione}, {Darmigny}, {de Boer}, {de Teodoro}, {de
  Zeeuw}, {Delle Luche}, {Domingues}, {Dubath}, {Fodor}, {Fr{\'e}zouls},
  {Fries}, {Fustes}, {Fyfe}, {Gallardo}, {Gallegos}, {Gardiol}, {Gebran},
  {Gomboc}, {G{\'o}mez}, {Grux}, {Gueguen}, {Heyrovsky}, {Hoar}, {Iannicola},
  {Isasi Parache}, {Janotto}, {Joliet}, {Jonckheere}, {Keil}, {Kim},
  {Klagyivik}, {Klar}, {Knude}, {Kochukhov}, {Kolka}, {Kos}, {Kutka}, {Lainey},
  {LeBouquin}, {Liu}, {Loreggia}, {Makarov}, {Marseille}, {Martayan},
  {Martinez-Rubi}, {Massart}, {Meynadier}, {Mignot}, {Munari}, {Nguyen},
  {Nordlander}, {Ocvirk}, {O'Flaherty}, {Olias Sanz}, {Ortiz}, {Osorio},
  {Oszkiewicz}, {Ouzounis}, {Palmer}, {Park}, {Pasquato}, {Peltzer}, {Peralta},
  {P{\'e}turaud}, {Pieniluoma}, {Pigozzi}, {Poels}, {Prat}, {Prod'homme},
  {Raison}, {Rebordao}, {Risquez}, {Rocca-Volmerange}, {Rosen}, {Ruiz-Fuertes},
  {Russo}, {Sembay}, {Serraller Vizcaino}, {Short}, {Siebert}, {Silva},
  {Sinachopoulos}, {Slezak}, {Soffel}, {Sosnowska}, {Strai{\v{z}}ys}, {ter
  Linden}, {Terrell}, {Theil}, {Tiede}, {Troisi}, {Tsalmantza}, {Tur},
  {Vaccari}, {Vachier}, {Valles}, {Van Hamme}, {Veltz}, {Virtanen}, {Wallut},
  {Wichmann}, {Wilkinson}, {Ziaeepour}, \& {Zschocke}}]{2016A&A...595A...1G}
{Gaia Collaboration}, {Prusti}, T., {de Bruijne}, J.~H.~J., {et~al.} 2016,
  \aap, 595, A1, \dodoi{10.1051/0004-6361/201629272}

\bibitem[{{Gaia Collaboration} {et~al.}(2022){Gaia Collaboration}, {Vallenari},
  {Brown}, {Prusti}, {de Bruijne}, {Arenou}, {Babusiaux}, {Biermann},
  {Creevey}, {Ducourant}, {Evans}, {Eyer}, {Guerra}, {Hutton}, {Jordi},
  {Klioner}, {Lammers}, {Lindegren}, {Luri}, {Mignard}, {Panem}, {Pourbaix},
  {Randich}, {Sartoretti}, {Soubiran}, {Tanga}, {Walton}, {Bailer-Jones},
  {Bastian}, {Drimmel}, {Jansen}, {Katz}, {Lattanzi}, {van Leeuwen}, {Bakker},
  {Cacciari}, {Casta{\~n}eda}, {De Angeli}, {Fabricius}, {Fouesneau},
  {Fr{\'e}mat}, {Galluccio}, {Guerrier}, {Heiter}, {Masana}, {Messineo},
  {Mowlavi}, {Nicolas}, {Nienartowicz}, {Pailler}, {Panuzzo}, {Riclet}, {Roux},
  {Seabroke}, {Sordo{\o}rcit}, {Th{\'e}venin}, {Gracia-Abril}, {Portell},
  {Teyssier}, {Altmann}, {Andrae}, {Audard}, {Bellas-Velidis}, {Benson},
  {Berthier}, {Blomme}, {Burgess}, {Busonero}, {Busso}, {C{\'a}novas}, {Carry},
  {Cellino}, {Cheek}, {Clementini}, {Damerdji}, {Davidson}, {de Teodoro},
  {Nu{\~n}ez Campos}, {Delchambre}, {Dell'Oro}, {Esquej},
  {Fern{\'a}ndez-Hern{\'a}ndez}, {Fraile}, {Garabato}, {Garc{\'\i}a-Lario},
  {Gosset}, {Haigron}, {Halbwachs}, {Hambly}, {Harrison}, {Hern{\'a}ndez},
  {Hestroffer}, {Hodgkin}, {Holl}, {Jan{\ss}en}, {Jevardat de Fombelle},
  {Jordan}, {Krone-Martins}, {Lanzafame}, {L{\"o}ffler}, {Marchal}, {Marrese},
  {Moitinho}, {Muinonen}, {Osborne}, {Pancino}, {Pauwels}, {Recio-Blanco},
  {Reyl{\'e}}, {Riello}, {Rimoldini}, {Roegiers}, {Rybizki}, {Sarro}, {Siopis},
  {Smith}, {Sozzetti}, {Utrilla}, {van Leeuwen}, {Abbas}, {{\'A}brah{\'a}m},
  {Abreu Aramburu}, {Aerts}, {Aguado}, {Ajaj}, {Aldea-Montero}, {Altavilla},
  {{\'A}lvarez}, {Alves}, {Anders}, {Anderson}, {Anglada Varela}, {Antoja},
  {Baines}, {Baker}, {Balaguer-N{\'u}{\~n}ez}, {Balbinot}, {Balog}, {Barache},
  {Barbato}, {Barros}, {Barstow}, {Bartolom{\'e}}, {Bassilana}, {Bauchet},
  {Becciani}, {Bellazzini}, {Berihuete}, {Bernet}, {Bertone}, {Bianchi},
  {Binnenfeld}, {Blanco-Cuaresma}, {Blazere}, {Boch}, {Bombrun}, {Bossini},
  {Bouquillon}, {Bragaglia}, {Bramante}, {Breedt}, {Bressan}, {Brouillet},
  {Brugaletta}, {Bucciarelli}, {Burlacu}, {Butkevich}, {Buzzi}, {Caffau},
  {Cancelliere}, {Cantat-Gaudin}, {Carballo}, {Carlucci}, {Carnerero},
  {Carrasco}, {Casamiquela}, {Castellani}, {Castro-Ginard}, {Chaoul},
  {Charlot}, {Chemin}, {Chiaramida}, {Chiavassa}, {Chornay}, {Comoretto},
  {Contursi}, {Cooper}, {Cornez}, {Cowell}, {Crifo}, {Cropper}, {Crosta},
  {Crowley}, {Dafonte}, {Dapergolas}, {David}, {David}, {de Laverny}, {De
  Luise}, {De March}, {De Ridder}, {de Souza}, {de Torres}, {del Peloso}, {del
  Pozo}, {Delbo}, {Delgado}, {Delisle}, {Demouchy}, {Dharmawardena}, {Di
  Matteo}, {Diakite}, {Diener}, {Distefano}, {Dolding}, {Edvardsson}, {Enke},
  {Fabre}, {Fabrizio}, {Faigler}, {Fedorets}, {Fernique}, {Fienga}, {Figueras},
  {Fournier}, {Fouron}, {Fragkoudi}, {Gai}, {Garcia-Gutierrez},
  {Garcia-Reinaldos}, {Garc{\'\i}a-Torres}, {Garofalo}, {Gavel}, {Gavras},
  {Gerlach}, {Geyer}, {Giacobbe}, {Gilmore}, {Girona}, {Giuffrida}, {Gomel},
  {Gomez}, {Gonz{\'a}lez-N{\'u}{\~n}ez}, {Gonz{\'a}lez-Santamar{\'\i}a},
  {Gonz{\'a}lez-Vidal}, {Granvik}, {Guillout}, {Guiraud},
  {Guti{\'e}rrez-S{\'a}nchez}, {Guy}, {Hatzidimitriou}, {Hauser}, {Haywood},
  {Helmer}, {Helmi}, {Sarmiento}, {Hidalgo}, {Hilger}, {H{\l}adczuk}, {Hobbs},
  {Holland}, {Huckle}, {Jardine}, {Jasniewicz}, {Jean-Antoine Piccolo},
  {Jim{\'e}nez-Arranz}, {Jorissen}, {Juaristi Campillo}, {Julbe}, {Karbevska},
  {Kervella}, {Khanna}, {Kontizas}, {Kordopatis}, {Korn}, {K{\'o}sp{\'a}l},
  {Kostrzewa-Rutkowska}, {Kruszy{\'n}ska}, {Kun}, {Laizeau}, {Lambert},
  {Lanza}, {Lasne}, {Le Campion}, {Lebreton}, {Lebzelter}, {Leccia}, {Leclerc},
  {Lecoeur-Taibi}, {Liao}, {Licata}, {Lindstr{\o}m}, {Lister}, {Livanou},
  {Lobel}, {Lorca}, {Loup}, {Madrero Pardo}, {Magdaleno Romeo}, {Managau},
  {Mann}, {Manteiga}, {Marchant}, {Marconi}, {Marcos}, {Marcos Santos},
  {Mar{\'\i}n Pina}, {Marinoni}, {Marocco}, {Marshall}, {Polo},
  {Mart{\'\i}n-Fleitas}, {Marton}, {Mary}, {Masip}, {Massari},
  {Mastrobuono-Battisti}, {Mazeh}, {McMillan}, {Messina}, {Michalik}, {Millar},
  {Mints}, {Molina}, {Molinaro}, {Moln{\'a}r}, {Monari}, {Mongui{\'o}},
  {Montegriffo}, {Montero}, {Mor}, {Mora}, {Morbidelli}, {Morel}, {Morris},
  {Muraveva}, {Murphy}, {Musella}, {Nagy}, {Noval}, {Oca{\~n}a}, {Ogden},
  {Ordenovic}, {Osinde}, {Pagani}, {Pagano}, {Palaversa}, {Palicio},
  {Pallas-Quintela}, {Panahi}, {Payne-Wardenaar}, {Pe{\~n}alosa Esteller},
  {Penttil{\"a}}, {Pichon}, {Piersimoni}, {Pineau}, {Plachy}, {Plum}, {Poggio},
  {Pr{\v{s}}a}, {Pulone}, {Racero}, {Ragaini}, {Rainer}, {Raiteri}, {Rambaux},
  {Ramos}, {Ramos-Lerate}, {Re Fiorentin}, {Regibo}, {Richards}, {Rios Diaz},
  {Ripepi}, {Riva}, {Rix}, {Rixon}, {Robichon}, {Robin}, {Robin}, {Roelens},
  {Rogues}, {Rohrbasser}, {Romero-G{\'o}mez}, {Rowell}, {Royer}, {Ruz Mieres},
  {Rybicki}, {Sadowski}, {S{\'a}ez N{\'u}{\~n}ez}, {Sagrist{\`a} Sell{\'e}s},
  {Sahlmann}, {Salguero}, {Samaras}, {Sanchez Gimenez}, {Sanna},
  {Santove{\~n}a}, {Sarasso}, {Schultheis}, {Sciacca}, {Segol}, {Segovia},
  {S{\'e}gransan}, {Semeux}, {Shahaf}, {Siddiqui}, {Siebert}, {Siltala},
  {Silvelo}, {Slezak}, {Slezak}, {Smart}, {Snaith}, {Solano}, {Solitro},
  {Souami}, {Souchay}, {Spagna}, {Spina}, {Spoto}, {Steele},
  {Steidelm{\"u}ller}, {Stephenson}, {S{\"u}veges}, {Surdej}, {Szabados},
  {Szegedi-Elek}, {Taris}, {Taylo}, {Teixeira}, {Tolomei}, {Tonello}, {Torra},
  {Torra}, {Torralba Elipe}, {Trabucchi}, {Tsounis}, {Turon}, {Ulla}, {Unger},
  {Vaillant}, {van Dillen}, {van Reeven}, {Vanel}, {Vecchiato}, {Viala},
  {Vicente}, {Voutsinas}, {Weiler}, {Wevers}, {Wyrzykowski}, {Yoldas}, {Yvard},
  {Zhao}, {Zorec}, {Zucker}, \& {Zwitter}}]{2022arXiv220800211G}
{Gaia Collaboration}, {Vallenari}, A., {Brown}, A.~G.~A., {et~al.} 2022, arXiv
  e-prints, arXiv:2208.00211, \dodoi{10.48550/arXiv.2208.00211}

\bibitem[{{Garc{\'\i}a} \& {Ballot}(2019)}]{2019LRSP...16....4G}
{Garc{\'\i}a}, R.~A., \& {Ballot}, J. 2019, Living Reviews in Solar Physics,
  16, 4, \dodoi{10.1007/s41116-019-0020-1}

\bibitem[{{Giammichele} {et~al.}(2018){Giammichele}, {Charpinet}, {Fontaine},
  {Brassard}, {Green}, {Van Grootel}, {Bergeron}, {Zong}, \&
  {Dupret}}]{2018Natur.554...73G}
{Giammichele}, N., {Charpinet}, S., {Fontaine}, G., {et~al.} 2018, \nat, 554,
  73, \dodoi{10.1038/nature25136}

\bibitem[{{Gough}(1993)}]{1993afd..conf..399G}
{Gough}, D.~O. 1993, in Astrophysical Fluid Dynamics - Les Houches 1987,
  399--560

\bibitem[{{Gough} \& {Thompson}(1991)}]{1991sia..book..519G}
{Gough}, D.~O., \& {Thompson}, M.~J. 1991, in Solar Interior and Atmosphere,
  519--561

\bibitem[{{Grevesse} \& {Sauval}(1998)}]{1998SSRv...85..161G}
{Grevesse}, N., \& {Sauval}, A.~J. 1998, \ssr, 85, 161,
  \dodoi{10.1023/A:1005161325181}

\bibitem[{{Gruberbauer} {et~al.}(2012){Gruberbauer}, {Guenther}, \&
  {Kallinger}}]{2012ApJ...749..109G}
{Gruberbauer}, M., {Guenther}, D.~B., \& {Kallinger}, T. 2012, \apj, 749, 109,
  \dodoi{10.1088/0004-637X/749/2/109}

\bibitem[{{Gruberbauer} {et~al.}(2013){Gruberbauer}, {Guenther}, {MacLeod}, \&
  {Kallinger}}]{2013MNRAS.435..242G}
{Gruberbauer}, M., {Guenther}, D.~B., {MacLeod}, K., \& {Kallinger}, T. 2013,
  \mnras, 435, 242, \dodoi{10.1093/mnras/stt1289}

\bibitem[{{Guo} \& {Jiang}(2023)}]{2023A&C....4200686G}
{Guo}, Z., \& {Jiang}, C. 2023, Astronomy and Computing, 42, 100686,
  \dodoi{10.1016/j.ascom.2023.100686}

\bibitem[{{Hon} {et~al.}(2020){Hon}, {Bellinger}, {Hekker}, {Stello}, \&
  {Kuszlewicz}}]{2020MNRAS.499.2445H}
{Hon}, M., {Bellinger}, E.~P., {Hekker}, S., {Stello}, D., \& {Kuszlewicz},
  J.~S. 2020, \mnras, 499, 2445, \dodoi{10.1093/mnras/staa2853}

\bibitem[{{Iglesias} \& {Rogers}(1993)}]{Iglesias1993}
{Iglesias}, C.~A., \& {Rogers}, F.~J. 1993, \apj, 412, 752,
  \dodoi{10.1086/172958}

\bibitem[{{Iglesias} \& {Rogers}(1996)}]{Iglesias1996}
---. 1996, \apj, 464, 943, \dodoi{10.1086/177381}

\bibitem[{{Irwin}(2004)}]{Irwin2004}
{Irwin}, A.~W. 2004, The FreeEOS Code for Calculating the Equation of State for
  Stellar Interiors.
\newblock \url{http://freeeos.sourceforge.net/}

\bibitem[{{Itoh} {et~al.}(1996){Itoh}, {Hayashi}, {Nishikawa}, \&
  {Kohyama}}]{Itoh1996}
{Itoh}, N., {Hayashi}, H., {Nishikawa}, A., \& {Kohyama}, Y. 1996, \apjs, 102,
  411, \dodoi{10.1086/192264}

\bibitem[{{Jermyn} {et~al.}(2021){Jermyn}, {Schwab}, {Bauer}, {Timmes}, \&
  {Potekhin}}]{Jermyn2021}
{Jermyn}, A.~S., {Schwab}, J., {Bauer}, E., {Timmes}, F.~X., \& {Potekhin},
  A.~Y. 2021, \apj, 913, 72, \dodoi{10.3847/1538-4357/abf48e}

\bibitem[{{Jermyn} {et~al.}(2022){Jermyn}, {Bauer}, {Schwab}, {Farmer}, {Ball},
  {Bellinger}, {Dotter}, {Joyce}, {Marchant}, {Mombarg}, {Wolf}, {Wong},
  {Cinquegrana}, {Farrell}, {Smolec}, {Thoul}, {Cantiello}, {Herwig}, {Toloza},
  {Bildsten}, {Townsend}, \& {Timmes}}]{Jermyn2022}
{Jermyn}, A.~S., {Bauer}, E.~B., {Schwab}, J., {et~al.} 2022, arXiv e-prints,
  arXiv:2208.03651.
\newblock \doarXiv{2208.03651}

\bibitem[{{Jiang} \& {Gizon}(2021)}]{2021RAA....21..226J}
{Jiang}, C., \& {Gizon}, L. 2021, Research in Astronomy and Astrophysics, 21,
  226, \dodoi{10.1088/1674-4527/21/9/226}

\bibitem[{{Kosovichev}(1999)}]{1999JCoAM.109....1K}
{Kosovichev}, A.~G. 1999, Journal of Computational and Applied Mathematics,
  109, 1

\bibitem[{{Kosovichev}(2011)}]{2011LNP...832....3K}
---. 2011, in Lecture Notes in Physics, Berlin Springer Verlag, ed. J.-P.
  {Rozelot} \& C.~{Neiner}, Vol. 832, 3, \dodoi{10.1007/978-3-642-19928-8_1}

\bibitem[{{Kosovichev} \& {Kitiashvili}(2020)}]{2020IAUS..354..107K}
{Kosovichev}, A.~G., \& {Kitiashvili}, I.~N. 2020, in Solar and Stellar
  Magnetic Fields: Origins and Manifestations, ed. A.~{Kosovichev},
  S.~{Strassmeier}, \& M.~{Jardine}, Vol. 354, 107--115,
  \dodoi{10.1017/S1743921320001416}

\bibitem[{{Kuhfuss}(1986)}]{1986A&A...160..116K}
{Kuhfuss}, R. 1986, \aap, 160, 116

\bibitem[{{Langanke} \& {Mart{\'{\i}}nez-Pinedo}(2000)}]{Langanke2000}
{Langanke}, K., \& {Mart{\'{\i}}nez-Pinedo}, G. 2000, Nuclear Physics A, 673,
  481, \dodoi{10.1016/S0375-9474(00)00131-7}

\bibitem[{{Lund} {et~al.}(2017){Lund}, {Silva Aguirre}, {Davies}, {Chaplin},
  {Christensen-Dalsgaard}, {Houdek}, {White}, {Bedding}, {Ball}, {Huber},
  {Antia}, {Lebreton}, {Latham}, {Handberg}, {Verma}, {Basu}, {Casagrande},
  {Justesen}, {Kjeldsen}, \& {Mosumgaard}}]{2017ApJ...835..172L}
{Lund}, M.~N., {Silva Aguirre}, V., {Davies}, G.~R., {et~al.} 2017, \apj, 835,
  172, \dodoi{10.3847/1538-4357/835/2/172}

\bibitem[{{Metcalfe} \& {Charbonneau}(2003)}]{2003JCoPh.185..176M}
{Metcalfe}, T.~S., \& {Charbonneau}, P. 2003, Journal of Computational Physics,
  185, 176, \dodoi{10.1016/S0021-9991(02)00053-0}

\bibitem[{{Metcalfe} {et~al.}(2009){Metcalfe}, {Creevey}, \&
  {Christensen-Dalsgaard}}]{2009ApJ...699..373M}
{Metcalfe}, T.~S., {Creevey}, O.~L., \& {Christensen-Dalsgaard}, J. 2009, \apj,
  699, 373, \dodoi{10.1088/0004-637X/699/1/373}

\bibitem[{{Metcalfe} {et~al.}(2023){Metcalfe}, {Townsend}, \&
  {Ball}}]{2023RNAAS...7..164M}
{Metcalfe}, T.~S., {Townsend}, R. H.~D., \& {Ball}, W.~H. 2023, Research Notes
  of the American Astronomical Society, 7, 164,
  \dodoi{10.3847/2515-5172/acebef}

\bibitem[{{Metcalfe} {et~al.}(2014){Metcalfe}, {Creevey}, {Do{\u{g}}an},
  {Mathur}, {Xu}, {Bedding}, {Chaplin}, {Christensen-Dalsgaard}, {Karoff},
  {Trampedach}, {Benomar}, {Brown}, {Buzasi}, {Campante}, {{\c{C}}elik},
  {Cunha}, {Davies}, {Deheuvels}, {Derekas}, {Di Mauro}, {Garc{\'\i}a},
  {Guzik}, {Howe}, {MacGregor}, {Mazumdar}, {Montalb{\'a}n}, {Monteiro},
  {Salabert}, {Serenelli}, {Stello}, {Ste\&{\c{s}}acute}, {licki}, {Suran},
  {Y{\i}ld{\i}z}, {Aksoy}, {Elsworth}, {Gruberbauer}, {Guenther}, {Lebreton},
  {Molaverdikhani}, {Pricopi}, {Simoniello}, \& {White}}]{2014ApJS..214...27M}
{Metcalfe}, T.~S., {Creevey}, O.~L., {Do{\u{g}}an}, G., {et~al.} 2014, \apjs,
  214, 27, \dodoi{10.1088/0067-0049/214/2/27}

\bibitem[{{Miglio} \& {Montalb{\'a}n}(2005)}]{2005A&A...441..615M}
{Miglio}, A., \& {Montalb{\'a}n}, J. 2005, \aap, 441, 615,
  \dodoi{10.1051/0004-6361:20052988}

\bibitem[{{Oda} {et~al.}(1994){Oda}, {Hino}, {Muto}, {Takahara}, \&
  {Sato}}]{Oda1994}
{Oda}, T., {Hino}, M., {Muto}, K., {Takahara}, M., \& {Sato}, K. 1994, Atomic
  Data and Nuclear Data Tables, 56, 231, \dodoi{10.1006/adnd.1994.1007}

\bibitem[{{Paxton} {et~al.}(2011){Paxton}, {Bildsten}, {Dotter}, {Herwig},
  {Lesaffre}, \& {Timmes}}]{Paxton2011}
{Paxton}, B., {Bildsten}, L., {Dotter}, A., {et~al.} 2011, \apjs, 192, 3,
  \dodoi{10.1088/0067-0049/192/1/3}

\bibitem[{{Paxton} {et~al.}(2013){Paxton}, {Cantiello}, {Arras}, {Bildsten},
  {Brown}, {Dotter}, {Mankovich}, {Montgomery}, {Stello}, {Timmes}, \&
  {Townsend}}]{Paxton2013}
{Paxton}, B., {Cantiello}, M., {Arras}, P., {et~al.} 2013, \apjs, 208, 4,
  \dodoi{10.1088/0067-0049/208/1/4}

\bibitem[{{Paxton} {et~al.}(2015){Paxton}, {Marchant}, {Schwab}, {Bauer},
  {Bildsten}, {Cantiello}, {Dessart}, {Farmer}, {Hu}, {Langer}, {Townsend},
  {Townsley}, \& {Timmes}}]{Paxton2015}
{Paxton}, B., {Marchant}, P., {Schwab}, J., {et~al.} 2015, \apjs, 220, 15,
  \dodoi{10.1088/0067-0049/220/1/15}

\bibitem[{{Paxton} {et~al.}(2018){Paxton}, {Schwab}, {Bauer}, {Bildsten},
  {Blinnikov}, {Duffell}, {Farmer}, {Goldberg}, {Marchant}, {Sorokina},
  {Thoul}, {Townsend}, \& {Timmes}}]{Paxton2018}
{Paxton}, B., {Schwab}, J., {Bauer}, E.~B., {et~al.} 2018, \apjs, 234, 34,
  \dodoi{10.3847/1538-4365/aaa5a8}

\bibitem[{{Paxton} {et~al.}(2019){Paxton}, {Smolec}, {Schwab}, {Gautschy},
  {Bildsten}, {Cantiello}, {Dotter}, {Farmer}, {Goldberg}, {Jermyn}, {Kanbur},
  {Marchant}, {Thoul}, {Townsend}, {Wolf}, {Zhang}, \& {Timmes}}]{Paxton2019}
{Paxton}, B., {Smolec}, R., {Schwab}, J., {et~al.} 2019, \apjs, 243, 10,
  \dodoi{10.3847/1538-4365/ab2241}

\bibitem[{{Pijpers}(2006)}]{2006mha..book.....P}
{Pijpers}, F.~P. 2006, {Methods in helio- and asteroseismology}

\bibitem[{{Pijpers} \& {Thompson}(1992)}]{1992A&A...262L..33P}
{Pijpers}, F.~P., \& {Thompson}, M.~J. 1992, \aap, 262, L33

\bibitem[{{Pijpers} \& {Thompson}(1994)}]{1994A&A...281..231P}
---. 1994, \aap, 281, 231

\bibitem[{{Rendle} {et~al.}(2019){Rendle}, {Buldgen}, {Miglio}, {Reese},
  {Noels}, {Davies}, {Campante}, {Chaplin}, {Lund}, {Kuszlewicz}, {Scott},
  {Scuflaire}, {Ball}, {Smetana}, \& {Nsamba}}]{2019MNRAS.484..771R}
{Rendle}, B.~M., {Buldgen}, G., {Miglio}, A., {et~al.} 2019, \mnras, 484, 771,
  \dodoi{10.1093/mnras/stz031}

\bibitem[{{Rogers} \& {Nayfonov}(2002)}]{Rogers2002}
{Rogers}, F.~J., \& {Nayfonov}, A. 2002, \apj, 576, 1064,
  \dodoi{10.1086/341894}

\bibitem[{{Roxburgh}(2017)}]{2017A&A...604A..42R}
{Roxburgh}, I.~W. 2017, \aap, 604, A42, \dodoi{10.1051/0004-6361/201731057}

\bibitem[{{Roxburgh} \& {Vorontsov}(2003)}]{2003A&A...411..215R}
{Roxburgh}, I.~W., \& {Vorontsov}, S.~V. 2003, \aap, 411, 215,
  \dodoi{10.1051/0004-6361:20031318}

\bibitem[{Santos {et~al.}(2018)Santos, Campante, Chaplin, Cunha, Lund, Kiefer,
  Salabert, García, Davies, Elsworth, \& Howe}]{Santos_2018}
Santos, A. R.~G., Campante, T.~L., Chaplin, W.~J., {et~al.} 2018, The
  Astrophysical Journal Supplement Series, 237, 17,
  \dodoi{10.3847/1538-4365/aac9b6}

\bibitem[{{Saumon} {et~al.}(1995){Saumon}, {Chabrier}, \& {van
  Horn}}]{Saumon1995}
{Saumon}, D., {Chabrier}, G., \& {van Horn}, H.~M. 1995, \apjs, 99, 713,
  \dodoi{10.1086/192204}

\bibitem[{{Silva Aguirre} {et~al.}(2015){Silva Aguirre}, {Davies}, {Basu},
  {Christensen-Dalsgaard}, {Creevey}, {Metcalfe}, {Bedding}, {Casagrande},
  {Handberg}, {Lund}, {Nissen}, {Chaplin}, {Huber}, {Serenelli}, {Stello}, {Van
  Eylen}, {Campante}, {Elsworth}, {Gilliland}, {Hekker}, {Karoff}, {Kawaler},
  {Kjeldsen}, \& {Lundkvist}}]{2015MNRAS.452.2127S}
{Silva Aguirre}, V., {Davies}, G.~R., {Basu}, S., {et~al.} 2015, \mnras, 452,
  2127, \dodoi{10.1093/mnras/stv1388}

\bibitem[{{Silva Aguirre} {et~al.}(2017){Silva Aguirre}, {Lund}, {Antia},
  {Ball}, {Basu}, {Christensen-Dalsgaard}, {Lebreton}, {Reese}, {Verma},
  {Casagrande}, {Justesen}, {Mosumgaard}, {Chaplin}, {Bedding}, {Davies},
  {Handberg}, {Houdek}, {Huber}, {Kjeldsen}, {Latham}, {White}, {Coelho},
  {Miglio}, \& {Rendle}}]{2017ApJ...835..173S}
{Silva Aguirre}, V., {Lund}, M.~N., {Antia}, H.~M., {et~al.} 2017, \apj, 835,
  173, \dodoi{10.3847/1538-4357/835/2/173}

\bibitem[{Sobol'(1967)}]{SOBOL196786}
Sobol', I. 1967, USSR Computational Mathematics and Mathematical Physics, 7,
  86, \dodoi{https://doi.org/10.1016/0041-5553(67)90144-9}

\bibitem[{{Teixeira} {et~al.}(2003){Teixeira}, {Christensen-Dalsgaard},
  {Carrier}, {Aerts}, {Frandsen}, {Stello}, {Maas}, {Burnet}, {Bruntt}, {de
  Medeiros}, {Bouchy}, {Kjeldsen}, \& {Pijpers}}]{2003Ap&SS.284..233T}
{Teixeira}, T.~C., {Christensen-Dalsgaard}, J., {Carrier}, F., {et~al.} 2003,
  \apss, 284, 233

\bibitem[{{Thompson} \& {Christensen-Dalsgaard}(2002)}]{2002ESASP.485...95T}
{Thompson}, M.~J., \& {Christensen-Dalsgaard}, J. 2002, in ESA Special
  Publication, Vol. 485, Stellar Structure and Habitable Planet Finding, ed.
  B.~{Battrick}, F.~{Favata}, I.~W. {Roxburgh}, \& D.~{Galadi}, 95--101.
\newblock \doarXiv{astro-ph/0110447}

\bibitem[{{Townsend} {et~al.}(2018){Townsend}, {Goldstein}, \&
  {Zweibel}}]{Townsend2018}
{Townsend}, R.~H.~D., {Goldstein}, J., \& {Zweibel}, E.~G. 2018, \mnras, 475,
  879, \dodoi{10.1093/mnras/stx3142}

\bibitem[{{Townsend} \& {Teitler}(2013)}]{Townsend2013}
{Townsend}, R.~H.~D., \& {Teitler}, S.~A. 2013, \mnras, 435, 3406,
  \dodoi{10.1093/mnras/stt1533}

\bibitem[{{Vandakurov}(1967)}]{1967AZh....44..786V}
{Vandakurov}, Y.~V. 1967, \azh, 44, 786

\bibitem[{{Vanlaer} {et~al.}(2023){Vanlaer}, {Aerts}, {Bellinger}, \&
  {Christensen-Dalsgaard}}]{2023A&A...675A..17V}
{Vanlaer}, V., {Aerts}, C., {Bellinger}, E.~P., \& {Christensen-Dalsgaard}, J.
  2023, \aap, 675, A17, \dodoi{10.1051/0004-6361/202245597}

\end{thebibliography}
\bibliographystyle{aasjournal}

\end{document}